\documentclass[10pt,journal,twocolumn,romanappendices]{IEEEtran}
\pdfoutput=1
\IEEEoverridecommandlockouts

\usepackage{algorithm,algorithmic,amsmath,amssymb,amsthm,array,bbm,cite,color,comment,graphicx,microtype,setspace,url,caption}

\usepackage[USenglish]{babel}
\usepackage{hyperref}
\usepackage[utf8]{inputenc}
\usepackage[T1]{fontenc}
\usepackage{enumitem}
\usepackage{graphicx} % For including images
\usepackage{subcaption} % For subfigures
\setitemize{itemsep=0mm,topsep=0mm,parsep=0pt,partopsep=0pt}

\usepackage{tikz}
\usetikzlibrary{arrows,backgrounds,calc,intersections,decorations.pathreplacing,positioning,shapes}

\makeatletter
\newcommand\subparagraph{%
  \@startsection{subparagraph}{5}
  {\parindent}
  {3.25ex \@plus 1ex \@minus .2ex}
  {-1em}
  {\normalfont\normalsize\bfseries}}
\makeatother

\usepackage{titlesec}
\let\subparagraph\relax
\titlespacing{\section}{3pt}{4pt plus 2pt minus 1pt}{3pt plus 2pt minus 1pt}
\titlespacing{\subsection}{3pt}{3pt plus 1pt minus 0pt}{2pt plus 1pt minus 0pt}
\usepackage{todonotes}
\usepackage[font=small]{caption}
\usepackage{comment}
\usepackage{epstopdf}
\usepackage{pgfplots}
\usepgfplotslibrary{groupplots}
\pgfplotsset{compat=1.17}
\newcommand{\h}{\mathbf{h}}

\newcommand{\m}{\mathbf{m}}

\renewcommand{\u}{\mathbf{u}}

\newcommand{\x}{\mathbf{x}}

\newcommand{\z}{\mathbf{z}}

\newcommand{\D}{\mathbf{D}}

\newcommand{\F}{\mathbf{F}}

\renewcommand{\H}{\mathbf{H}}
\newcommand{\I}{\mathbf{I}}

\newcommand{\Y}{\mathbf{Y}}
\newcommand{\Z}{\mathbf{Z}}

\newcommand{\setA}{\mathcal{A}}

\newcommand{\setC}{\mathcal{C}}
\newcommand{\setD}{\mathcal{D}}

\newcommand{\setG}{\mathcal{G}}

\newcommand{\setI}{\mathcal{I}}

\newcommand{\setM}{\mathcal{M}}
\newcommand{\setN}{\mathcal{N}}

\newcommand{\Compl}{\mbox{$\mathbb{C}$}}
\newcommand{\rmF}{\mathrm{F}}

\newcommand{\blkdiag}{\mathrm{blkdiag}}
\newcommand{\diag}{\mathrm{diag}}

\newcommand{\Exp}{\mathbb{E}}
\newcommand{\herm}{\mathrm{H}}
\renewcommand{\Im}{\mathrm{Im}}

\renewcommand{\Re}{\mathrm{Re}}

\newcommand{\tran}{\mathrm{T}}

\title{Near-Field MIMO Channel Acquisition: Geometry-Aided Feedback and Transmission Design}
\author{Shima Eslami,~\IEEEmembership{Student Member,~IEEE}, Bikshapathi Gouda,~\IEEEmembership{Member,~IEEE}, \\ and Antti Tölli,~\IEEEmembership{Senior Member,~IEEE}
\thanks{The authors are with the Centre for Wireless Communications, University of Oulu, Finland (e-mail: \{shima.eslami, bikshapathi.gouda, antti.tolli\}@oulu.fi). This work is supported by the Research Council of Finland under grant no. 24303893 (CAMAIDE). %Part of this work was presented at \color{red} Previous conference 
This work builds on our previous studies presented at IEEE ICASSP 2024~\cite{myIcassp}, and IEEE SAM 2024~\cite{SAM}.
\color{black}}}
\begin{document}

\maketitle

\begin{abstract}
Near-field (NF) line-of-sight (LoS) MIMO systems enable efficient channel state information (CSI) acquisition and precoding by exploiting known antenna geometries at both the base station (BS) and user equipment (UE). This paper introduces a compact parameterization of the NF LoS MIMO channel using two angles of departure (AoDs) and a BS-UE relative rotation angle. The inclusion of the second AoD removes the need for fine-grained distance grids imposed by conventional NF channel parametrization.
To address the user-specific uplink pilot overhead in multiuser NF CSI acquisition, we propose a scheme that uses a fixed, UE-independent set of downlink pilots transmitted from a carefully selected subset of BS antennas. In dominant LoS conditions, as few as four pilots suffice, with Cramér-Rao bound (CRB) analysis confirming that increased antenna spacing improves estimation accuracy. Each UE estimates and quantizes its angular parameters and feeds them back to the BS for geometry-based CSI reconstruction, eliminating the need for full channel feedback.
To enhance robustness against noise, quantization errors, and non-line-of-sight (NLoS) components, we introduce a two-stage precoding method. The initial precoding is computed from estimated LoS CSI and refined through bidirectional over-the-air (OTA) training. Furthermore, a two-step stream allocation strategy reduces pilot and computational overhead. Simulations demonstrate that the proposed approach achieves high data rates with significantly fewer OTA iterations, approaching the performance of perfect CSI.

\end{abstract}
%*******************************
\section{Introduction}
\label{section:1}
%My Research Plan
The transition to beyond  5G networks involves increased demands for higher data rates, improved energy efficiency, reduced latency, and support for a more diverse range of users compared to 5G~\cite{Pennanen}.
At higher frequencies, the radiating near-field range expands significantly due to the larger apertures of antenna arrays operating at these frequencies~\cite{Dai_Drnzo_review}. Frequencies in the FR3 range (7–24 GHz), as well as millimeter wave (mmWave) and subterahertz (THz) bands, exemplify this phenomenon by offering extensive bandwidth and allowing large-scale highly directional antenna systems~\cite{Pennanen, Zhang_Eldar, Nir}. This expansion suggests that in future 6G networks, communication may increasingly occur in the near-field (NF) scenarios rather than being confined to the far-field (FF)~\cite{Nir,cui_Dai2022near}.

The boundary between the NF and FF regions is defined by the Rayleigh distance. Beyond this distance, electromagnetic fields can be approximated as planar waves, while within it, NF propagation dominates and must be modeled using spherical waves~\cite{cui_Dai2022near}.
The spherical wavefront in the NF offers both opportunities and challenges for system design and signal processing~\cite{Unloking_potentials}. 
It enables more focused beams that can direct signals to specific spatial points, thereby enhancing interference management~\cite{Nir,cui_Dai2022near} and improving spectral efficiency in multiuser scenarios~\cite{Dai_Multipleaccess,pherical_SEff}.
In contrast, traditional FF beam steering is limited to pointing beams in a specific direction~\cite{Dai_Drnzo_review}.
Moreover, NF line-of-sight (LoS) propagation can boost the spatial multiplexing gain of MIMO systems~\cite{Dai_Drnzo_review}. However, models and techniques developed for FF conditions are often inadequate, requiring novel approaches to fully exploit the potential of NF communication~\cite{Nir, FF_NF_different}.

Accurate channel state information (CSI) acquisition at the base station (BS) is essential for effective beam focusing~\cite{Localization}. Unlike FF channels, which rely primarily on the angle of arrival/departure (AoA/AoD) and exhibit angular-domain sparsity, NF channels also depend on the distance between transmit (TX) and receive (RX) antennas~\cite{Dai_NForFF}. This dual dependency necessitates revising sparse representation models and estimation algorithms~\cite{Eldar2}.
To capture both angle and distance information, a polar-domain representation was introduced in~\cite{Dai_NForFF}, enabling compressed sensing (CS)-based channel estimation with low pilot overhead. To further reduce complexity, \cite{Eldar2} proposes a distance-parameterized angular-domain sparse model, where the dictionary size depends only on angular resolution. Channel estimation, including both LoS and non-line-of-sight (NLoS) components, is performed via joint dictionary learning and sparse recovery, treating the user distance as an unknown dictionary parameter.
For wideband NF channels, \cite{Dai_Wideband} introduces a CS-based algorithm that estimates AoA and distance while addressing the beam-split effect by exploiting its bilinear structure. In~\cite{Dai_NForFF, Eldar2, Dai_Wideband}, both LoS and NLoS channels are reconstructed using CS-based methods.

Unlike FF LoS channels, which typically have a rank-one structure, NF channels can exhibit higher ranks, enabling spatial multiplexing~\cite{Jarkko3D, Jarkko-spawc, Nitin}. The NF LoS MIMO channel can be efficiently reconstructed using a few geometric parameters, such as the TX-RX antenna-pair-specific reference distance, AoA(AoD), and relative transceiver rotation~\cite{Dai_mixedLoS, Jarkko3D}. This property allows for accurate channel estimation with minimal pilot overhead~\cite{Jarkko3D, Jarkko-spawc, Nitin, Dai_mixedLoS, Localization, ch_est2_ar}.
In~\cite{Jarkko3D, Jarkko-spawc, Nitin}, the NF LoS channel is decomposed into parallel subchannels, where an FF approximation is applied to each separately. Using geometric relationships among subchannel-specific FF AoAs, a likelihood function is computed for the estimation of the LoS parameters using a message passing algorithm. Similarly,~\cite{Dai_mixedLoS} represents the NF MIMO channel using three geometric parameters where an initial maximum likelihood (ML) estimation is refined via a gradient-based algorithm, while NLoS components are recovered using a CS-based method.
%For more structured parameter estimation,
\cite{Localization} models LoS, reflection, and NLoS paths of an NF MISO channel using three geometric parameters for a uniform planar array (UPA) at the BS, estimating them via a damped Newtonized OMP algorithm. In~\cite{ch_est2_ar}, a sequential approach decouples angle and distance estimation, first applying a 2-dimensional discrete Fourier transform for angle estimation, followed by a low-complexity closed-form distance estimation method.

Most existing NF parameter estimation and sparse channel recovery methods rely on dedicated uplink pilots per user, leading to significant overhead and potential pilot contamination as user numbers grow~\cite{Downlink}.%~\cite{Downlink,Pilot_Contamination}. 
Additionally, uplink-based estimation assumes full channel reciprocity and precise RF calibration, which may be impractical. Conventional downlink-based CSI acquisition, on the other hand, suffers from excessive feedback overhead~\cite{fdd}.  
To address these challenges, we propose a CSI acquisition framework leveraging the parametric representation of the NF LoS channel. This approach uses a minimal number of downlink pilots and limited geometry information exchange between network nodes. Instead of acquiring the full LoS MIMO channel, only 
%three angular parameters are estimated: arrival, departure, and rotation for a selected pair of reference antennas. 
three angular parameters are estimated: relative TX-RX rotation and two AoDs from a carefully selected pair of reference antennas. The second AoD is introduced to address distance sampling challenges in existing works.
Users estimate these parameters based on known antenna placements and a few pilots from a subset of BS antennas. The quantized estimates are then fed back to the BS, enabling full LoS MIMO channel reconstruction with minimal overhead.  
However, estimation noise, quantization errors, and unaccounted NLoS components can degrade the reconstructed channel and impact the precoding/combining design. If strong NLoS components are overlooked, relying solely on LoS estimates may be ineffective. To mitigate this, we incorporate bidirectional over-the-air (OTA) beamformer training, eliminating the need for explicit NLoS estimation. Unlike traditional OTA training, which starts with random precoding/combining~\cite{BDT,Biksh,Antti_BiT,Praneeth}, our approach exploits the acquired multiuser LoS CSI at the BS for efficient initialization, significantly accelerating convergence.  
The main contributions of our work are summarized as follows:
\begin{itemize}
    \item Minimal pilot overhead in CSI acquisition: We propose a multiuser NF LoS MIMO CSI acquisition scheme where the BS broadcasts a fixed, user-independent set of downlink pilots. Unlike conventional methods that require pilots from all BS antennas, our scheme transmits only from a subset of antennas. For instance, when LoS dominates, as few as \(4 \ll M\) pilots suffice, with $M$ as the number of the BS antennas. We show that maximized spacing between selected pilot-transmitting antennas is crucial.
  
    \item Geometry-based channel reconstruction: Each user estimates three geometric parameters using full knowledge of its inter-antenna distances and partial knowledge of BS antenna spacing. Quantized estimates are fed back via in-band or out-of-band signaling, eliminating full MIMO channel feedback. Given the known antenna placements, the BS reconstructs the LoS channels of the users using the geometric relationships between the TX and RX antennas.

    \item Efficient channel parameterization: 
    The NF LoS MIMO channel is defined by three angular parameters (two AoDs and a rotation angle) along with known antenna spacings to infer all TX-RX distances. The second AoD replaces the conventional reference distance, eliminating grid challenges and enabling uniform angle sampling for simpler estimation. We analyze the impact of pilot count, array size, and relative TX-RX orientation on estimation accuracy using Cramér-Rao Lower Bound (CRLB) analysis.

    \item Two-phase precoding/combining design: In the first phase, the BS computes initial precoding solely based on the reconstructed LoS CSI using a symmetric rate design criterion. In the second phase, an OTA iterative training process refines the precoding (and combining) to account for NLoS effects. Our approach, initialized with geometry-based LoS estimates, converges significantly faster than random initialization and approaches the performance of perfect CSI with minimal OTA iterations.

    \item Optimized stream allocation: To further accelerate convergence and minimize pilot overhead, we implement a two-step stream allocation process. First, streams are assigned based on the rank of the estimated LoS channels. In the second step, allocations are refined based on the rates achieved using first-phase precoding.
\end{itemize}
This work builds on earlier conference papers~\cite{myIcassp,SAM}, which addressed LoS channel reconstruction with minimal downlink pilots and feedback, and OTA-based precoding refinement under non-negligible NLoS components, respectively. Here, we extend these results by analyzing LoS CSI acquisition accuracy using the CRLB, the likelihood function, and a symmetric rate (i.e., max-min user rate) criterion. Furthermore, we examine the influence of system parameters—array length, pilot counts, and relative TX-RX orientation—and introduce improved reference angle preselection and stream allocation to reduce design complexity and overhead.
\noindent The paper is structured as follows: Section~\ref{sec:SM} discusses the system model, including NF LoS MIMO channel model and its reference parameters.  Section~\ref{sec:ch-est} covers LoS CSI acquisition, parameter estimation, and feedback. 
Section~\ref{crlb}
 provides CRLB derivation. Section~\ref{sec:Precoder_Design} presents the proposed two-stage precoding/combining design. Simulation results are in Section~\ref{sec:num}, and conclusions in Section~\ref{sec:CONC}.
% The remainder of this paper is structured as follows: Section~\ref{sec:SM} introduces the %multiuser system and near-field channel model, 
% system model, discussing the NF LoS channel parameters and channel estimation using uplink and downlink pilots. Section~\ref{sec:ch-est} presents the LoS CSI acquisition procedure with downlink pilots, describing the parameter estimation, feedback, and channel recovery. The CRLB derivation is described. In Section~\ref{sec:Precoder_Design}, we detail our proposed two-stage precoder/combiner design procedure. Simulation results are provided in Section~\ref{sec:num}, and conclusions are drawn in Section~\ref{sec:CONC}.

%=============================================
\textit{Notation:}
%=============================================
\color{black}
For a matrix $\mathbf{X}$, $\mathbf{X}^\tran$, $\mathbf{X}^\herm$, and $\mathbf{X}^{-1}$ denote the transpose, conjugate transpose, and inverse, respectively. The entry in the $n^\text{th}$ row and $m^\text{th}$ column of $\mathbf{X}$ is $\mathbf{X}_{[n,m]}$, and the $m^\text{th}$ column is $\mathbf{X}_{[:,m]}$. The real and imaginary parts of $\mathbf{X}$ are $\Re[\mathbf{X}]$ and $\Im[\mathbf{X}]$, respectively.
$\blkdiag(\mathbf{X}_1,\cdots,\mathbf{X}_N)$ constructs a block diagonal matrix from $\mathbf{X}_1,\cdots,\mathbf{X}_N$, while $\diag({x}_1,\cdots,{x}_N)$ creates a diagonal matrix with elements of ${x}_1,\cdots,{x}_N$ on the diagonal. Horizontal concatenation of vectors $\mathbf{x}_n$ is written as $[\mathbf{x}_1,\cdots,\mathbf{x}_N]$, and $\{ \mathbf{x}_{n} \}_{n=1}^N$ denote the set of vectors $\mathbf{x}_n$. For a set $\setM$, $\setM(i)$ is its $i^\text{th}$ member, and $|\setM|$ is its cardinality. The Euclidean and Frobenius norms are $\| \cdot \|$ and $\| \cdot \|_{\rmF}$, respectively.
The gradient with respect to $x$ is $\nabla_{x}(\cdot)$, and $\mathcal{L}_{(\textrm{P})}$ denotes the Lagrangian of optimization problem $(\textrm{P})$. The notation $\mathbf{F}(\x)$ or $\mathbf{F}(\x_{[1]},\cdots,\x_{[M]})$ indicates that each element of $\mathbf{F}$ is a function of components of $\x$. %The circularly symmetric complex Gaussian distribution is $\setC \setN (\mathbf{m}, \mathbf{C})$, with $\mathbf{m}$ as the mean and $\mathbf{C}$ as the covariance matrix. Finally, $\mathbf{1}$ is a vector of ones, $\textsf{j}=\sqrt{-1}$, and $\setI_N={1,\cdots,N}$.

\color{black}
 
%--------------------------
\section{System Model}
\label{sec:SM}
%\subsection{Multiuser MIMO setup}
%\label{multiuser}
We consider a downlink multiuser MIMO system where a single BS with $M$ antennas serves $K$ user equipments (UEs), with each UE~$k$ equipped with $N_k$ antennas and allocated $S_k$ data streams.
%We consider a multiuser MIMO setup with a single base station (BS) equipped with $M$ antennas to serve $K$ user equipments (UEs), where UE~$k$ equipped with $N_k$ antennas and allocated with $S_k$ data streams in the downlink. %we assume a time division duplex (TDD) system with channel reciprocity. Furthermore, for simplicity, we consider a narrowband channel.%\footnote{Extending this framework to incorporate the frequency selectivity of channels is an interesting direction for future research.} 
Let $\H_{k} \in \Compl^{N_k \times M}$ be the downlink channel between the BS and the UE $k$, and $x_{k,s} \sim \setC \setN (0,1)$ be the data stream~$s$ of UE $k$. Let $\m_{k,s}\in \Compl^{M \times 1}$ be the precoding vector designed by the BS to serve the data stream $s$ of UE $k$, and then the signal received by UE $k$ is expressed as
%********************************
\begin{align} \label{eq:y_k}
 {\mathbf{y}}_{k}=  \sum_{j=1}^{K} \sum_{s =1}^{S_{j}}\H_{k} 
 \m_{j,s}  x_{j,s} + \z_{k} \in \Compl^{N_k \times 1},
\end{align}
%**********************************
where $\z_k \sim \setC \setN (\mathbf{0},\sigma^2_k \I_{N_k})$, and $S_{j}$ is the number of streams of the UE $j$.  
Considering the linear receiver combining, the signal-to-interference-plus-noise ratio (SINR) of stream~$s$ of UE~$k$ is given as
%---------------------------
% \begin{align} \label{eq:SINR_sk}
% \Gamma_{k,s}= \frac{ |\u_{k,s}^{\herm} \H_{k} \m_{k,s}|^{2}}{  \sum_{\tilde k=1}^K \sum_{\tilde s=1}^{S_{\tilde k}}   |\u_{k,s}^{\herm} \H_{k} \m_{\tilde k,\tilde s} |^{2} + \sigma_{k}^{2} \| \u_{k,s} \|^{2}},
% \end{align}
\begin{align} \label{eq:SINR_sk}
\Gamma_{k,s} = \frac{ |\u_{k,s}^{\herm} \H_{k} \m_{k,s}|^{2}}{  %\sum\limits_{\tilde k=1}^K 
\sum\limits_{\substack{%\tilde s=1 \\ 
(j, l) \neq (k, s)}}%^{S_{\tilde k}}
|\u_{k,s}^{\herm} \H_{k} \m_{j,l} |^{2} + \sigma_{k}^{2} \| \u_{k,s} \|^{2}},
\end{align}
%---------------------------
where $\u_{k,s}\in\Compl^{N_k \times 1}$ is the combining vector at UE~$k$ for $x_{k,s}$. 
%It is apparent from \eqref{eq:SINR_sk} that the design of precoders depends on the combiner vectors and vice-versa. In section \color{red} ref \color{black} we discuss the precoder/combiner design procedure in more detail. 
Finally, the rate of UE $k$ (in bps/Hz) is given by
\begin{align} \label{eq:R_k}
R_k= \sum_{s=1}^{S_k} \log_{2}(1 + \Gamma_{k,s}).
\end{align}
In general, each UE-specific channel can be decomposed into sum of the LoS channel $\H_k^{\text{LoS}}$ and the NLoS channel $\H_k^{\text{NLoS}}$ as $\H_k= \H_k^{\text{LoS}}+\H_k^{\text{NLoS}}$. In the following section, we describe the LoS MIMO channel between the BS and each UE, highlighting the key channel parameters under the NF assumption.
% In general, each user-specific channel can be decomposed to the LoS and NLoS components as $\H_k= \H_k^{\text{LoS}}+\H_k^{\text{NLoS}}$.
%In the following, we discuss the channel model in more detail.
%*****************************
\subsection{LoS MIMO Channel Model and NF channel parameters}
 %Since this paper mainly focuses on leveraging the properties of the LoS channel for transmission design, we only discuss the LoS component and leave the NLoS channel description for Section~\ref{sec:num}. 
For ease of notation, from this section until Section~\ref{sec:Precoder_Design}, we consider an arbitrary UE and omit the user index 
$k$ from the channel and other associated parameters. %In Section~\ref{sec:Precoder_Design}, we reintroduce the index $k$ to distinguish between user-specific symbols.
% we omit the user index $k$ in the following and will reintroduce it in section~\ref{sec:Precoder_Design}. 
Without loss of generality, we assume that both the BS and the UE are equipped with linear antenna arrays with arbitrary antenna spacing. Moreover, both the BS and the UEs are assumed to have a fully digital architecture.\footnote{The framework can also be extended to other antenna architectures, such as (non-)uniform planar arrays, and also hybrid transceiver structures.} 
%We also assume that the BS and user's antenna arrays have the same polarization and are coplanar. 
Now we describe the LoS channel component.
%using a geometric model. 
Let $d_{n,m}$ denote the distance between the $m^\text{th}$ BS antenna %($m\in \setI_M$) 
and the $n^\text{th}$ UE antenna elements. %($n\in \setI_N$). 
The $(n,m)^\text{th}$ entry of the LoS channel matrix is given by \cite{Indoor_mmw}
%-----------------------
\begin{align} \label{eq:H_LoS}
\H^{\text{LoS}}_{[n,m]} = \frac{\lambda}{4\pi d_{n,m}}
\:\textit{exp}\:\bigg(\frac{-\textsf{j}2\pi d_{n,m}} {\lambda}\bigg),
\end{align}
%------------------------
where $\lambda$ is the carrier wavelength. 
%-------------------------
\begin{figure}[t!]
    \centering   \includegraphics[scale=0.3]{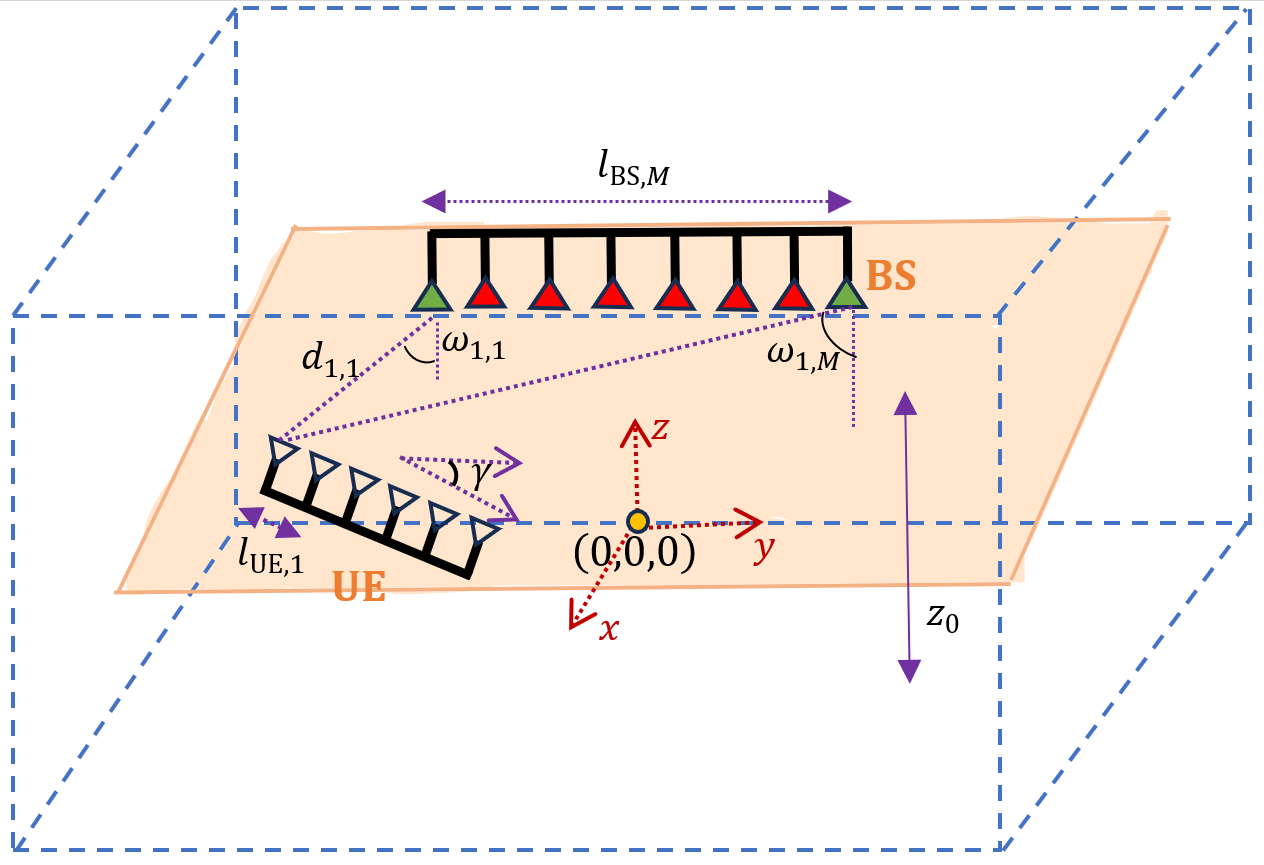} % Adjust scale as needed
    \caption{LoS channel parametrization}
 %   \vspace{-0.5cm}
    \label{fig:system_model1}
\end{figure}
%-------------------------
To explore the geometric properties of the channel, we define antenna pair-specific AoDs $\omega_{n,m}$ as the angle 
%AoD/AoA %made by the ray 
% between the 
% $m^\text{th}$ antenna element at the BS and the $n^\text{th}$ UE antenna element. 
%between $d_{n,m}$ and the normal of the BS.
between the BS broadside direction and the departure direction from the $m^\text{th}$ BS antenna towards the $n^\text{th}$ UE antenna.  
We also define $\gamma$, as the rotation angle of the UE relative to the direction of the BS array, as shown in Figure~\ref{fig:system_model1}.
Then, the antenna-pair-specific distances $d_{n,m}$ can be represented as \eqref{eq:r_nm}, shown on the top of the next page, where $l_{\text{BS},m}$ is the distance between the $m^\text{th}$ and first antenna element at the BS, and similarly,  $l_{\text{UE},n}$ represents the antenna spacing between the first and the $n^\text{th}$ UE antenna.\footnote{Geometric parameters are generally defined in a 3D coordinate system. To simplify the presentation, and without loss of generality, we assume all antenna arrays are confined to a common horizontal plane, thereby restricting parameter estimation to a 2D subspace.}
%***********************
\begin{figure*}[t!]
%\addtocounter{equation}{+0}
\begin{align}\label{eq:r_nm}
d_{n,m} = \sqrt{ d_{1,1}^2+2d_{1,1}\Big(l_{\text{BS},m} \sin\omega_{1,1}-l_{\text{UE},n} \sin(\omega_{1,1}+\gamma)\Big)-2l_{\text{BS},m}l_{\text{UE},n} \cos \gamma+l_{\text{BS},m}^2+l_{\text{UE},n}^2},
\end{align}
\hrulefill
\end{figure*}
%************************
In addition, the antenna pair-specific AoDs are obtained as
%------------------
\begin{align}\label{eq:theta_nm}
\omega_{n,m}=\tan ^{-1} \bigg(\frac{d_{1,1} \sin\omega_{1,1}-l_{\text{UE},n}\cos\gamma+l_{\text{BS},m}}{d_{1,1}\cos\omega_{1,1}-l_{\text{UE},n}\sin\gamma}\bigg).
\end{align}
%------------------
It can be seen from \eqref{eq:r_nm} that the antenna-pair-specific distances and consequently the LoS channel matrix can be represented based on three geometric parameters, i.e., $d_{1,1}$, $\omega_{1,1}$ (relative to a reference antenna pair), and $\gamma$ (capturing the relative TX-RX rotation). In other words, $\H^\text{LoS}$ can be fully characterized by the triplet $(d_{1,1},\omega_{1,1},\gamma)$ assuming that the antenna spacing parameters are known in advance. 
The geometric interpretation of this observation is that the UE (BS) can determine the
%the location (relative to the UE orientation) of the first BS antenna 
 position of the first BS (UE) antenna relative to its own orientation using $d_{1,1}$ and $\omega_{1,1}$. 
 The relative locations of the remaining BS (UE) antennas can be found on the basis of the first BS (UE) antenna position, the rotation angle $\gamma$, and the known BS (UE) antenna spacing.
Thus, the NF LoS channel estimation problem is reduced to a geometric parameter estimation problem.

%\color{gray}It should be noted that for simplicity, we have assumed that both the BS and user arrays are located within the same horizontal plane, as depicted in Fig .~\ref {fig:system_model1}. 
 %This arrangement can be visualized in an environment where the BS is mounted on a sidewall at a height $z_0$, while the user is positioned within a horizontal plane at $z=z_0$. An example, as discussed in~\cite{Jarkko-spawc}, could involve the BS array being installed on an augmented reality headset and the user array on a wearable device such as a smartwatch. However, in practice, the BS can be located on a different plane with varying heights. Consider, for example, a large auditorium where the BS antennas are distributed along the ceiling, while users, e.g., laptops, are randomly placed on a horizontal plane at $z=z_0$. This scenario is further detailed in Appendix~\ref{app:A1}.\color{black}
%**********************
%***************************
%\subsection{NF LoS MIMO Channel Parametrization}
%
%\subsection{Reference distance challenges, introducing Angular parameter}
%******************************
\subsection{Conversion of Distance-Angle Grid to Joint Angular Grid}
Estimating the NF channel $\H^\text{LoS}$ reduces to determining three reference parameters: $(d_{1,1}, \omega_{1,1}, \gamma)$. Specifically, 
%the NF LoS channel triplet can be obtained by searching over a three-dimensional (3D) grid to find the optimal value of a utility function defined by that triplet. To do this, the 3D joint angle-distance grid $\tilde\setA$ can be defined as
The NF LoS channel triplet can be obtained by searching over a joint angle-distance grid to find the optimal value of a utility function defined by that triplet. The associated three-parameter search space, $\boldsymbol{\setG}$, can be defined as:
%Specifically, the NF LoS channel triplet can be obtained by minimizing the squared error between the measured signal and its approximation. This involves searching for the reference parameters over a 3-dimensional(3D) joint angle-distance grid $\tilde\setA$
% The NF  $\H^\text{LoS}$ is equivalent to estimating three reference parameters i.e., $(d_{1,1},\omega_{1,1},\gamma)$. 
% In other words, the NF LoS channel's triplet can be estimated by e.g., minimizing the squared error between the measured signal and its approximation, where the reference distance and angles are searched through a 3D discrete angle-distance joint grid $\tilde\setA$ to detect the reference parameters, where
%
% To achieve this, a maximum likelihood (ML) problem is typically formulated to minimize the %energy of the difference
% squared error between the measured signal and its approximation, where the reference distance and angles are searched through a 3D discrete angle-distance joint grid $\tilde\setA$ to detect the reference parameters, where
%-----------------
\begin{align}
  \boldsymbol{\setG}=\{(d_{1,1},\omega_{1,1},\gamma); (\omega_{1,1},\gamma) \in \boldsymbol{\Omega},d_{1,1}\in \boldsymbol{\setD} ) \},  
\end{align}
%-----------------
where $\boldsymbol{\Omega}$ is the two-parameter angle search space and $\boldsymbol{\setD}$ is the distance grid.
To construct the $\boldsymbol{\Omega}$ grid, the angles are typically sampled uniformly, i.e.,
%-----------------
\begin{align}
  &\boldsymbol{\Omega}(\omega_{1,1},\gamma) = \nonumber \\
  &\Bigg\{ \bigg( 
  \frac{\omega_\text{min}(L_\omega - i) + i\omega_\text{max}}{L_\omega}, 
  \frac{\gamma_\text{min}(L_\gamma - j) + j\gamma_\text{max}}{L_\gamma}
  \bigg) 
  \Bigg\},  
\end{align}
%-----------------
for all $i\in\{0\}\cup\setI_{L_\omega}$ and $j\in\{0\}\cup\setI_{L_\gamma}$, where $L_\omega$, $L_\gamma$, are
the number of AoD and the rotation angle samples, respectively, and $\omega_{1,1}\in [\omega_\text{min},\omega_\text{max}]$ and $\gamma_{1,1}\in [\gamma_\text{min},\gamma_\text{max}]$. However, uniform distance sampling may prove to be inefficient as the distance parameter has a significant impact on the amplitude and phase variation of the signal in NF scenarios, while its impact diminishes as the distance increases~\cite{Localization}. In addition, prior knowledge of the distance range is required for uniform sampling.
To address this issue, one can adopt the two-step distance sampling approach of~\cite{Localization}, where an exponential distance sampling strategy is used at the first step in obtaining a coarse estimate $(\hat{d}_{1,1},\hat{\omega}_{1,1},\hat{\gamma})$ of the geometric parameters. This distance grid is defined in~\cite{Localization} as 
%-----------------
\begin{align}
  \boldsymbol{\setD}(d_{1,1}) = 
  \Big\{ (d_\text{Ray})^\frac{r}{L_d};\quad \forall r\in \setI_{L_d}
  \Big\},  
\end{align}
%-----------------
where $d_\text{Ray}$ is the Rayleigh distance and $L_d$ is number of distance samples. In the second phase, a hierarchical distance dictionary is utilized, where the distance is sampled uniformly around $\hat{d}_{1,1}$ and the coarse estimate of distance is updated with higher accuracy by fixing the angles and searching the distance using the fine distance grid. In particular, the estimator must know the dimensions of the arrays to calculate $d_\text{Ray}$. Also, in high-frequency setups where the wavelength is extremely small, the step size of distance sampling must be sufficiently fine. To overcome these challenges and avoid the second search, we introduce an alternative angle parameter $\omega_{n,m} \in [\omega_\text{min}, \omega_\text{max}]$, with $(n,m) \ne (1,1)$ rather than distance. We select $\omega_{1,M}$, the AoD at the last BS antenna to maximize its angular separation from $\omega_{1,1}$, thereby improving parameter distinguishability during estimation. Thus, from \eqref{eq:theta_nm}, we obtain
%-----------------
\begin{align} \label{eq:r11}
d_{1,1} = \frac{l_{\text{BS},M}} {\tan\omega_{1,M}\cos\omega_{1,1}-\sin\omega_{1,1}}.
\end{align}
%--------------------
Substituting \eqref{eq:r11} into \eqref{eq:r_nm} redefines the LoS channel parameters as $(\omega_{1,1},\omega_ {1,M},\gamma)$, which means that the intersection of two outermost AoDs can determine the location of the first antenna of the UE.
%Note that the second reference AoD $\omega_ {1,M}$ is selected at the last BS antenna to maximize its angular separation from $\omega_ {1,1}$, thereby enhancing the distinguishability between the two parameters during the estimation process.
Having replaced the reference distance with an angle, the joint angular grid $\boldsymbol{\setA}_1$ constructed using three uniformly sampled angles is introduced as
%-----------------
\begin{align}  &\boldsymbol{\setA}_1(\omega_{1,1},\omega_{1,M},\gamma)=\boldsymbol{\Omega}(\omega_{1,1},\gamma)\bigcup \nonumber \\  &\Bigg\{ \bigg( 
  \frac{\omega_\text{min}(L_\omega - i) + i\omega_\text{max}}{L_\omega}
  \bigg) 
  \Bigg\}, \quad \forall i\in\{0\}\cup \setI_{L_\omega}.
\end{align}
%-----------------
Finally, we remove the infeasible elements from $\boldsymbol{\setA}_1$, i.e., the elements that correspond to negative distances, and form the final angular grid $\boldsymbol{\setA}$ as 
% Finally, the suitable grid $\boldsymbol{\setA}$ for estimationproblem is obtained by removing the infeasible elements resulting in negative distances from $\boldsymbol{\setA}_1$ as
%---------------
\begin{align}
\boldsymbol{\setA}=\Big\{(\omega_{1,1},\omega_{1,M},\gamma)\in \boldsymbol{\setA}_1 , d_{1,1}{(\omega_{1,1},\omega_{1,M},\gamma)}\geq 0\Big \}.
\end{align}
\section{Geometry-aided Near-Field LoS MIMO CSI Acquisition from Downlink Pilots}
\label{sec:ch-est}
%In this section, we outline the entire LoS MIMO channel acquisition procedure based on downlink pilots for an arbitrary user. 
In this section, we outline the proposed NF LoS MIMO CSI acquisition procedure. We begin by comparing the LoS CSI acquisition using downlink and uplink pilots. Then, we describe the channel parameter estimation procedure at the UE using downlink pilots. Finally, we describe the CSI acquisition at the BS using the channel parameters obtained from the UE.
%\subsection{LoS CSI Acquisition: Downlink Pilot v.s. Uplink Pilots}

\begin{figure}[t]
    \centering
    \includegraphics[width= \columnwidth]{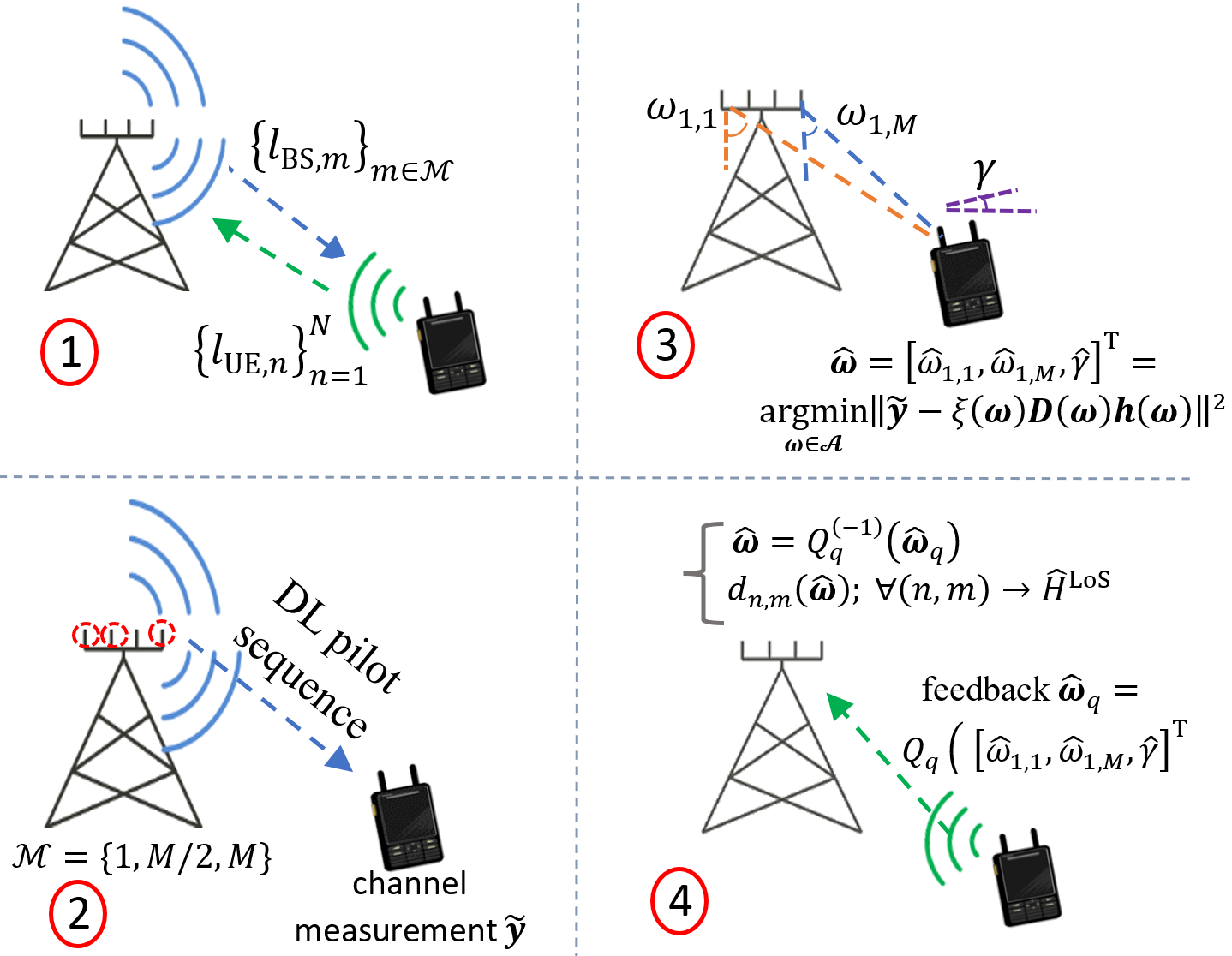}
    \caption{NF LoS MIMO CSI acquisition via downlink pilots. }
  % \vspace{-0.6cm}
    \label{fig:DL_est}
\end{figure}

%Downlink CSI at the BS can be acquired using either uplink or downlink pilots. To estimate the NF LoS multi-user MIMO channel from uplink pilots,
%the UEs transmit a set of orthogonal pilots in the uplink.
To acquire the downlink LoS MIMO CSI at the BS from user-specific uplink pilots, the BS must estimate the channel parameters
%, of the UEs 
$\{(\omega_{1,1},\omega_ {1,M},\gamma)_k\}_{k=1}^K$ and reconstruct the corresponding LoS MIMO channels $\{ \mathbf{H}^{\text{LoS}}_k\}_{k=1}^K$. This process requires LoS channel reciprocity and RF component calibration. Moreover, the pilot overhead increases linearly with the number of UEs, becoming significant as $K$ grows. 
%each UE transmits its pilots orthogonal in the uplink. 
% The BS estimates the channel parameters of the UEs $\{(\omega_{1,1},\omega_ {1,M},\gamma)_k\}_{k=1}^K$ for all $k\in \setI_K$ and reconstructs the MIMO channels $\{ \mathbf{H}_k\}_{k=1}^K$. In this setup, as the number of UEs increases, the pilot overhead increases significantly. In addition, full-channel reciprocity and RF component calibration are required.
On the other hand, in LoS CSI acquisition schemes based on downlink pilots, such as the one proposed in this paper, each UE estimates its channel parameters $(\omega_{1,1},\omega_ {1,M},\gamma)$ from a small set of (antenna-specific) pilots broadcast by the BS. The estimated parameters are then returned to the BS during the feedback phase in the uplink. As a result, the number of required pilots does not scale with the number of UEs. Moreover, returning only three geometric parameters per UE significantly reduces the feedback overhead compared to traditional downlink-based channel estimation methods, where the entire channel matrix must be fed back.
% in the downlink-based LoS channel acquisition scheme, such as the one proposed in this paper, the BS broadcasts the pilots to all UEs. Thus, the number of pilots does not scale with the number of UEs. Each UE estimates its parameters $(\omega_{1,1},\omega_ {1,M},\gamma)$ and sends them back to the BS during the feedback phase. 
% As a result, returning only three geometric parameters significantly reduces feedback overhead compared to traditional channel estimation methods using downlink pilots, where the entire channel must be fed back.
It is noted that UEs need to know $\{l_{\text{BS},m}\}_{m\in \setM}$, where $\setM$ is the set of indices of BS antennas used for pilot transmission. This information can be broadcast through a control channel before the pilot transmissions. After collecting the feedback parameters, the BS can reconstruct the LoS MIMO channels $\{\mathbf{H}_k^\text{LoS}\}_{k=1}^K$. Notably, for both uplink/downlink-based LoS CSI acquisition schemes, the antenna spacing parameters of the UEs, i.e., $\{\{(l_{\text{UE},n})_k\}_{n=1}^{N_k}\}_{k=1}^K$ must be known at the BS. These parameters must be communicated to the BS before the estimation begins. Figure~\ref{fig:DL_est} illustrates the procedure outlined above. It is worth noting that, as the BS reconstructs the channels, the scheme leverages the higher computational capability of the BS compared to that of the UE. 
%In this section, we outline our proposed NF LoS MIMO CSI acquisition procedure.We begin by comparing LoS CSI acquisition using downlink and uplink pilots. Then, we describe the pilot transmission and channel approximation models. Next, we detail the parameter estimation and channel recovery phases. Finally, we derive the CRLB for the estimation problem.
%********************************
%*******************************
\subsection{LoS Channel Parameter Estimation at the UE}
%Recall that we defined the set 
To describe the pilot transmission and reception process, we define $\setM$ to denote the set of indices of BS antennas active for downlink pilot transmission, where $|\setM|\geq 2$. We assume that $\setM(1)=1$ and $\setM(|\setM|)=M$, meaning that the two outermost BS antennas are always selected for pilot transmission to maximize spatial separation, with the remaining antenna indices selected in between. Our proposed method uses $|\setM|\ll M$ pilot sequence to recover the LoS MIMO channel.
To model the pilots, we define the matrix $\mathbf{P}\in \mathbb{C}^{|\setM|\times \tau}$, where $\tau$ is the length of each pilot sequence and $\mathbf{P}\mathbf{P}^\herm=\tau \mathbf{I}_{|\setM|}$.
Having transmitted the pilot sequences from the active BS antennas, the UE receives the signal ${\mathbf{Y}} \in \mathbb{C}^{N\times \tau}$ as
%.......................
\begin{align}
\label{Y_tilde}
{\mathbf{Y}} = \rho\Big[\mathbf{H}_{[:,\setM(1)]},\cdots, \mathbf{H}_{[:,\setM(|\setM|)]}\Big]\mathbf{P} + {\mathbf{Z}},  
\end{align}
%Another notation:
%\Big[\mathbf{H}\big(:,\setM(1)\big),\cdots, \mathbf{H}\big(:,\setM(|\setM|)\big)\Big]
%......................
where $\rho$ is the BS total transmit power for pilot transmission and ${\mathbf{Z}}$ is receiver noise such that each element follows the distribution of $\mathcal{CN} (0,\sigma^2)$. The UE correlates the received signal 
with the known pilot matrix to extract the channel measurement matrix $\tilde{\mathbf{Y}}\in \mathbb{C}^{N\times|\setM|}$ as 
%.......................
\begin{align}
\label{Y}
&\tilde{\mathbf{Y}}\triangleq \frac{1}{\tau}{\mathbf{Y}}\mathbf{P}^\herm=\rho\Big[\mathbf{H}_{[:,\setM(1)]},\cdots, \mathbf{H}_{[:,\setM(|\setM|)]}\Big] + \tilde{\mathbf{Z}},  
\end{align}
%......................
where $\tilde{\mathbf{Z}}\triangleq {\mathbf{Z}}\mathbf{P}^\herm/\tau$ which its elements follows the distribution $ \mathcal{CN} (0,\tilde\sigma^2)$, where $\tilde\sigma={\sigma}/\tau$. Now, let us consider the received signal from the $m^\text{th}$ BS antenna, which is given by
%.......................
\begin{align}
\label{ym}
&\tilde{\mathbf{y}}_m=\rho \mathbf{H}_{[:,m]}+ \tilde{\mathbf{z}}_m  \in \mathbb{C}^{N}; \quad m\in \setM
\end{align}
%......................
where $\tilde{\mathbf{y}}_m=\tilde{\mathbf{Y}}_{[:,m]}$, and %$\tilde{\mathbf{z}}_m\sim \mathcal{CN} (0,\tilde\sigma^2 \mathbf{I}_N)$. 
$\tilde{\mathbf{z}}_m=\tilde{\mathbf{Z}}_{[:,m]}$. 
Each channel $\mathbf{H}_{[:,m]}$ can be approximate as $\mathbf{H}_{[:,m]}\approx\mathbf{D}_m {\mathbf{h}}_m$, where
\begin{align}
\label{Gh}
    {\mathbf{h}}_m  \triangleq  \bigg[\textit{exp}\:\bigg(\frac{-\textsf{j}2\pi d_{{1,m}}} {\lambda}\bigg), \ldots, \textit{exp}\:\bigg(\frac{-\textsf{j}2\pi d_{{N,m}}} {\lambda}\bigg)\bigg]^{\tran},
\end{align}
and 
$\D_m=\diag(d_{1,m}^{-1},\cdots,d_{N,m}^{-1} )$. As mentioned in Section~\ref{sec:SM}, all distance parameters $d_{n,m}$ are a function of  $(\omega_{1,1},\omega_{1,M},\gamma)$. For ease of notation, let us put the LoS channel parameters in a vector and define $\boldsymbol{\omega}\triangleq [\omega_{1,1},\omega_{1,M},\gamma]^\tran$.
Thus, for all $m\in \setM$, the signal $\tilde{\mathbf{y}}_m$ can be approximated as
%------------------
\begin{align}
\label{ym_apprx2}
\tilde{\mathbf{y}}_m\approx {\xi} \mathbf{D}_m(\boldsymbol{\omega}){\mathbf{h}}_{m}(\boldsymbol{\omega})+ \tilde{\mathbf{z}}_m,
\end{align} 
%-------------------
Where $\xi$ is a complex-valued factor that compensates for mismatches between the received signal from all pilot-transmitting BS antennas and the approximation terms $\mathbf{D}_m\mathbf{h}_m, \forall m\in \setM$. %and must also be estimated.
% where $\xi$ is a
% complex-valued factor to compensate for the mismatches between the received signal from all pilot transmitting BS antennas and the approximation terms  $\mathbf{D}_m\mathbf{h}_m, \forall m\in \setM$, and needs to be estimated as well. 
%The closed-form expression for the estimate of $\xi_m$ is given by
To estimate the angular vector $\boldsymbol{\omega}$ and the correction factor $\xi$, we concatenate the received signal and noise vectors as $\tilde{\mathbf{y}}\triangleq\text{vec}(\tilde{\mathbf{Y}})=[\tilde{\mathbf{y}}_{\setM(1)}^\tran,\cdots,\tilde{\mathbf{y}}_{\setM(|\setM|)}^\tran ]^\tran$, and $\mathbf{z}\triangleq\text{vec}(\tilde{\mathbf{Z}})=[\tilde{\mathbf{z}}_{\setM(1)}^\tran,\cdots,\tilde{\mathbf{z}}_{\setM(|\setM|)}^\tran ]^\tran$. Thus,
\eqref{Y} can be approximated as
%-----------------
\begin{align}
\label{y}
\tilde{\mathbf{y}}\approx{{\xi}}\D(\boldsymbol{\omega}){\h}(\boldsymbol{\omega})+ \tilde{\mathbf{z}},
\end{align}
%-----------------
where $\D(.)=\blkdiag\Big(\D_{\setM(1)}(.),\cdots,\D_{\setM(|\setM|)}(.)\Big)$, and ${\boldsymbol{h}}(.)=\big[{\boldsymbol{h}}_{\setM(1)}^\tran(.),\cdots,{\boldsymbol{h}}_{\setM(|\setM|)}^\tran(.)\big]^{\tran}$.
Now $\xi$ can be obtained in closed form as
%---------------------------
\begin{align}
\label{alpha}
\xi &= \mathop{\text{argmin}}_{\substack{\xi\in \mathbb{C}}} \|\tilde{\mathbf{y}}- \xi\D(\boldsymbol{\omega}) {\h}(\boldsymbol{\omega}) \|^2= \frac{{\h}^\herm(\boldsymbol{\omega})\D(\boldsymbol{\omega})\tilde{\mathbf{y}}}{N\|\D(\boldsymbol{\omega})\|_\text{F}^2}.
\end{align}
We can now formulate the ML-based estimation of the LoS channel parameters on the UE side, which is given by
%---------------------------
\begin{equation}\label{eq:ML problem1} \hat{\boldsymbol{\omega}}=\mathop{\text{argmin}}_{\substack{\boldsymbol{\omega}\in \boldsymbol{\setA}}}  \Big\|\tilde{\mathbf{y}}-{\xi}(\boldsymbol{\omega})\D(\boldsymbol{\omega}) {\h}(\boldsymbol{\omega})\Big\|^2,
\end{equation}
%-------------------------
where $\hat{\boldsymbol{\omega}}=[\hat{\omega}_{1,1},\hat{\omega}_{1,M},\hat{\gamma}]^\tran$.
% It is worth noting that the correction factor $\xi_m$ can also be defined as common for all active BS antennas, i.e., $\xi_m = \xi, \:  \forall m\in \setM$. 
% In this case, 
%---------------------------
Substituting~\eqref{alpha} into~\eqref{eq:ML problem1}, yields
%---------------------------
\begin{equation}\label{eq:ML problem} \hat{\boldsymbol{\omega}}=\mathop{\text{argmin}}_{\substack{\boldsymbol{\omega}\in \boldsymbol{\setA}}}  \Big\|\tilde{\mathbf{y}}- \frac{{\h}^\herm(\boldsymbol{\omega})\D(\boldsymbol{\omega})\tilde{\mathbf{y}}}{N\|\D(\boldsymbol{\omega})\|_\text{F}^2}\D(\boldsymbol{\omega}) {\h}(\boldsymbol{\omega})\Big\|^2.
\end{equation}
%-------------------------
 In summary, to estimate the NF LoS MIMO channel parameters, each UE computes the squared error between the channel measurement signal $\tilde{\mathbf{y}}$ and its approximation, as in~\eqref{eq:ML problem1}, for each grid point $({\omega}_{1,1},{\omega}_{1,M},{\gamma}) \in \boldsymbol{\setA}$, and identifies the grid point that minimizes this squared error.
To compute $\mathbf{D}(\boldsymbol{\omega}){\mathbf{h}}(\boldsymbol{\omega})$, the UE must know the antenna spacing parameters for all its elements and the pilot-transmitting antennas at the BS.
In this regard, we assume that each UE is fully aware of its antenna array geometry, while the BS broadcasts the inter-antenna distances of its active antennas using  control channels, e.g., physical downlink control channel (PDCCH),
%or DCCH %\textbf{complete}..., 
before the pilot transmission phase.
%----------------------
%.......................
%\subsection{Channel Parameter Feedback and Reconstruction at the BS}
\subsection{Angle Estimate Feedback and LoS Channel Reconstruction}
 Upon completing the estimation of the LoS channel parameters, the feedback phase is initiated by quantizing the estimated angles at the UE. Let $Q_q(.)$ denote a $q$-bit quantization function, and $\hat{\boldsymbol{\omega}}_q=Q_q\left( \hat{\boldsymbol{\omega}}\right)$ represent the quantized angle estimate vector computed by the UE. The UE transmits $\hat{\boldsymbol{\omega}}_q$ to the BS via uplink, using either in-band or out-of-band channels.
 %The UE transmits $\hat{\boldsymbol{\omega}}_q$ to the BS via the uplink, where 
% the feedback transmission can occur through either in-band or out-of-band channels.
 %transmission can be carried out using either in-band or out-of-band channels.
 The BS collects feedback angles from the UE and extracts the angle estimates as $\hat{\boldsymbol{\omega}}=Q_q^{(-1)}\left( \hat{\boldsymbol{\omega}}_q\right)$, where $Q^{(-1)}_q(.)$ is dequantization function. Then it proceeds with the LoS channel reconstruction phase based on the elements of $\hat{\boldsymbol{\omega}}$. %i.e., $\left( \hat{\omega}_{1,1}, \hat{\omega}_{1,M},\hat{\gamma}\right)$. 
 To this end, the BS first calculates $\hat{d}_{1,1}$ based on the angles $( \hat{\omega}_{1,1}, \hat{\omega}_{1,M})$ from~~\eqref{eq:r11}, given that it has prior knowledge of its own antenna array length. It then utilizes $(\hat{d}_{1,1},\hat{\omega}_{1,1},\hat{\gamma})$ to estimate the antenna-pair-specific distances $\hat{d}_{n,m}$ from~\eqref{eq:r_nm}, leveraging the known antenna-spacing parameters of the transceivers, i.e., $\{l_{\text{UE},n}\}_{n=1}^{n=N}$, and $\{l_{\text{BS},m}\}_{m=1}^{n=M}$. 
 %To ensure the availability of these parameters at the BS, Accordingly, To make these parameters available at the BS
 To provide the BS with these parameters, we assume that the BS is aware of its array geometry and that each UE transmits its antenna spacing $\{l_{\text{UE},n}\}_{n=1}^{n=N}$ to the BS via control channels, e.g., physical uplink control channel (PUCCH), before the pilot transmission stage.  
Finally, the entries of the LoS channel matrix, $\mathbf{H}^\text{LoS}$, are recovered by substituting $\hat{d}_{n,m}$ into~\eqref{eq:H_LoS}. Figure~\ref{fig:DL_est} and Algorithm~\ref{alg1} illustrate the LoS CSI acquisition process at the BS, which is based on downlink pilots for an arbitrary UE. In a multiuser scenario, the procedure runs independently for each UE.
As a byproduct of channel reconstruction, the BS can estimate UE antenna positions relative to its coordinates 
%the relative positions of UE antennas with respect to its own coordinates 
using $(\hat{d}_{1,1}, \hat{\omega}_{1,1}, \hat{\gamma})$. Assuming the BS array lies along the y-axis with the origin at its center, and given the UE antenna spacings, the estimated UE coordinates in $x\text{-}y$ plane are
\begin{align}
    \hat{x}^\text{UE}_n &= \hat{x}^\text{UE}_1 - l_{\text{UE},n} \sin\hat{\gamma}, \nonumber \\
    \hat{y}^\text{UE}_n &= \hat{y}^\text{UE}_1 + l_{\text{UE},n} \cos\hat{\gamma}, \quad n=2,\dots,N,
\end{align}
where
$\hat{x}^\text{UE}_1 = \hat{x}^\text{BS}_1 + \hat{d}_{1,1} \cos\hat{\omega}_{1,1}$, and $\hat{y}^\text{UE}_1 = \hat{y}^\text{BS}_1 - \hat{d}_{1,1} \sin\hat{\omega}_{1,1}$.
%This provides the UE’s position relative to the BS coordinate system, enabling positioning applications.
%-------------------------------------------

\begin{algorithm}
%\begin{algorithmic}
%\color{gray}
  \textbf{Data:} Pilot matrix $\mathbf{P}$, the angle grid $\boldsymbol{\setA}$, $\{l_{\text{BS},m}\}_{m=1}^M$, $\{l_{\text{UE},n}\}_{n=1}^N$, and the set $\setM$.    
 % \hspace{8mm}$\{l_{\text{TX},m}\}_{j=1}^M$ and $\{l_{\text{RX},i}\}_{i=1}^N$, respectively.
 
 \begin{itemize}[leftmargin=12mm]
\item[\texttt{(S.0)}]
%****************************
%\begin{itemize}%[leftmargin=12mm]
\textbf{BS}: Broadcasts $\{l_{\text{BS},m}\}_{m\in\setM}$ via a control channel,\\ 
\textbf{UE}: Send $\{l_{\text{UE},n}\}_{n=1}^N$  via a control channel.
%\end{itemize}
%*****************************
\item[\texttt{(S.1)}]
%\begin{itemize}
\textbf{BS}: Send $|\setM|\ll M$  downlink pilots $\mathbf{P}$.  %from the  antenna set $\setM$.

\textbf{UE}: Collect the channel measurement $\mathbf{Y}$ (see~\eqref{Y_tilde}), correlate it with pilot as $\tilde{\mathbf{Y}}=\frac{1}{\tau}\mathbf{Y}\mathbf{P}^\herm$ and form $\tilde{\mathbf{y}}=\text{vec}(\tilde{\mathbf{Y}})$.
%\end{itemize}
%*****************************
\item[\texttt{(S.2)}]
%\begin{itemize}
 \textbf{UE}: Approximate $\tilde{\mathbf{y}}$ as ${\xi}(\boldsymbol{\omega}) \D (\boldsymbol{\omega}) {\h}(\boldsymbol{\omega})$ for each $\boldsymbol{\omega} \in \boldsymbol{\setA}$ from~\eqref{Gh}-\eqref{y}, %and obtain $\hat{\boldsymbol{\omega}}$ from~\eqref{eq:ML problem1}
  %\item[] \textbf{UE}: 
and estimate the LoS channel angular parameters as $\hat{\boldsymbol{\omega}}$ from~\eqref{eq:ML problem1}.
%\end{itemize}
%*****************************
\item[\texttt{(S.3)}] 
%\begin{itemize}
 \textbf{UE}: Quantize angle estimates as $\hat{\boldsymbol{\omega}}_q=Q_q(\hat{\boldsymbol{\omega}})$ and feedback them in the UL. 
%\end{itemize}
%*****************************
\item[\texttt{(S.4)}]
%\begin{itemize}
\textbf{BS}:  Extract $\hat{\boldsymbol{\omega}}=Q^{(-1)}_q(\hat{\boldsymbol{\omega}}_q)$ by dequantizing received feedback, and calculate $\hat{d}_{11}$ from~\eqref{eq:r11} based on $\hat{\boldsymbol{\omega}}$.%$(\hat{\boldsymbol{\omega}}_{[1]},\hat{\boldsymbol{\omega}}_{[2]})$.
% obtain $d_{1,1} ({\scriptstyle Q( \hat{\boldsymbol{\omega}})})$
% and $\big\{\big\{d_{n,m} ({\scriptstyle Q( \hat{\boldsymbol{\omega}})})\}_{n=1}^N\big\}_{m=1}^M$
%  as in~\eqref{eq:r11} and \eqref{eq:r_nm}.
\\
\textbf{BS}: Obtain $\{\{\hat{d}_{n,m}\}_{n=1}^N\}_{m=1}^M$ based on 
$(\hat{d}_{11}, \hat{\omega}_{1,1},\hat{\gamma})$ from~\eqref{eq:r_nm}, and
calculate the entries of $\hat{\mathbf{H}}^\text{LoS}$ as in \eqref{eq:H_LoS}.
%\end{itemize}
 \end{itemize}
 %\end{algorithmic}
\caption{NF LoS MIMO CSI acquisition at the BS based on the UE feedback (for each UE).} \label{alg1}
\end{algorithm}
%_____________________________________
%\vspace{-0.5cm}
%===============================
%******************************
\begin{figure*}[t]
%\hrule
\begin{align}
\addtocounter{equation}{+8}
    \frac{\partial d_{n,m}}{\partial \omega_{1,1}} &= \frac{d_{1,1}}{d_{n,m}} \bigg(\frac{\tan\omega_{1,M}\sin\omega_{1,1}+\cos\omega_{1,1}} {\tan\omega_{1,M}\cos\omega_{1,1}-\sin\omega_{1,1}} 
     \Big( d_{1,1} + l_{\text{BS},m} \sin\omega_{1,1} - l_{\text{UE},n} \sin(\omega_{1,1}+\gamma) \Big)
    \nonumber \\
    &
    \phantom{=}+  l_{\text{BS},m} \cos\omega_{1,1} - l_{\text{UE},n} \cos(\omega_{1,1}+\gamma) \bigg) \label{doh_d_doh_omega11} \\
     \frac{\partial d_{n,m}}{\partial \omega_{1,M}} &= \frac{d_{1,1}}{d_{n,m}} \bigg(\frac{-\sec^2\omega_{1,M}\cos\omega_{1,1}} {\tan\omega_{1,M}\cos\omega_{1,1}-\sin\omega_{1,1}} 
    \Big( d_{1,1} + l_{\text{BS},m} \sin\omega_{1,1} - l_{\text{UE},n} \sin(\omega_{1,1}+\gamma) \Big)\bigg) \label{doh_d_doh_omega1M}
\end{align}
\addtocounter{equation}{-10} 
\hrule
\end{figure*}
%*************************
%===============================
% \textcolor{magenta}{to be completed}
%===============================
\section{CRLB Analysis}
\label{crlb}
% IN this section, we derived the CRLB for the asymptotic estimate of $\theta_1,...\theta..$.... for this we define...
% according to CRLB.. the error variance of each estimate ....should less than crlB. which is invese of fisher information...
% first equation..
% To compute the FIM... we define ...$\mu$..
% considering $\tilde y$.....
% the $i,j$ the element of fim can be written as..
% write equatron..27..
% then define the variable here..
In this section, we derive the CRLB to evaluate the performance limits and the accuracy of our estimation problem given in~\eqref{eq:ML problem1}. 
We define the real-valued parameter vector $\boldsymbol{\varepsilon}$ as
%***********************
\begin{align} \label{eps} \boldsymbol{\varepsilon} \triangleq \big[\omega_{1,1},\omega_{1,M},\gamma,\Re[\xi], \Im[\xi] \big]^\tran. \end{align}
%**************************
The corresponding ML estimate is denoted by $\hat{\boldsymbol{\varepsilon}}$. Specifically, the subvector $\hat{\boldsymbol{\varepsilon}}_{[1:3]} = \hat{\boldsymbol{\omega}}$ is obtained from~\eqref{eq:ML problem}, and $\hat{\boldsymbol{\varepsilon}}_{[4:5]}$ is derived from~\eqref{alpha}. The estimate $\hat{\boldsymbol{\varepsilon}}$ is assumed to be asymptotically unbiased. Consequently, the CRLB for the mean squared error (MSE) of the $i^\text{th}$ element of $\hat{\boldsymbol{\varepsilon}}$ is given by
% Consider a real-valued parameter vector $\boldsymbol{\varepsilon}$ and its unbiased estimator $\hat{\boldsymbol{\varepsilon}}$. 
%*********************
\begin{align}
\label{mse_crb}
e_i \triangleq\Exp\Big\{\big| \hat{\boldsymbol{\varepsilon}}_{[i]}-\boldsymbol{\varepsilon}_{[i]}\big|^2\Big\}\geq \mathbf{C}_{[i,i]} , 
\end{align}
%*********************
where $\mathbf{C}$ is the inverse of the Fisher information matrix (FIM) $\mathbf{F}$, i.e., $\mathbf{C}^{-1}=\mathbf{F}$~\cite{kay1993fundamentals}. 
To calculate FIM for our problem described in \eqref{eq:ML problem1}, let us define 
    \begin{equation}
    \label{mu}
        \boldsymbol{\mu}(\boldsymbol{\varepsilon})=\boldsymbol{\mu}(\boldsymbol{\omega},\xi)\triangleq{\xi}\D(\boldsymbol{\omega}) {\h}(\boldsymbol{\omega})\in \mathbb{C}^{N|\setM|},
    \end{equation}
%$\boldsymbol{\mu}(\boldsymbol{\varepsilon})=\boldsymbol{\mu}(\boldsymbol{\omega},\xi)\triangleq{\xi}\D(\boldsymbol{\omega}) \tilde{\h}(\boldsymbol{\omega})\in \mathbb{C}^{N|\setM|}$,
where $\xi$ is treated as an independent estimation parameter. Thus, for a given $\boldsymbol{\omega}$ and $\xi$ and assuming LoS channel, we have $\tilde{\mathbf{y}} \sim \mathcal{CN} \big(\boldsymbol{\mu},\tilde\sigma^2 \mathbf{I}_N\big)$.  The $(i,j)^\text{th}$ element of the FIM corresponding to the white Gaussian observation is given by~\cite{CRLB}
%----------------------
\begin{equation}
\label{Fij}   \mathbf{F}_{[i,j]} = \Re\bigg[\frac{2}{\tilde\sigma^2}\frac{\partial \boldsymbol{\mu}^{\herm}(\boldsymbol{\varepsilon})}{\partial \boldsymbol{\varepsilon}_{[i]} } \frac{\partial \boldsymbol{\mu}(\boldsymbol{\varepsilon}) }{\partial \boldsymbol{\varepsilon}_{[j]} }\bigg].
\end{equation}
%------------------
% where the parameter vector for our problem is defined as 
% %---------------
% \begin{align}
% \label{eps}
% \boldsymbol{\varepsilon}=\big[\omega_{1,1},\omega_{1,M},\gamma,\Re[\xi],\Im[\xi]\big]^\tran
% \end{align}
%---------------------------------
For the computation of the FIM $\mathbf{F}\in \mathbb{R}^{5\times5}$, let us omit the arguments $(\boldsymbol{\varepsilon})$ and $(\boldsymbol{\omega})$ from the variables, e.g., 
$\boldsymbol{\mu}$, ${\mathbf{h}}$, and $\mathbf{D}$, in the following.
From $\boldsymbol{\mu}$ definition~\eqref{mu} we have 
%--------------------
\begin{equation}
\label{doh_mu_doh_xi}
\frac{\partial \boldsymbol{\mu}}{\partial {\Re[\xi]}}= \frac{-\textsf{j}\partial \boldsymbol{\mu}}{\partial {\Im[\xi]}}=\mathbf{D}{\mathbf{h}}.
\end{equation}
%--------------------------
Furthermore, for an angle $\omega \in \{\omega_{1,1}, \omega_{1,M}, \gamma\}$ we have
%-------------------------
\begin{equation}   \label{doh_mu_doh_omega1}
\frac{\partial \boldsymbol{\mu}}{\partial {\omega}}= \xi\left(\frac{\partial \boldsymbol{\mathbf{D}}}{\partial {\omega}}{\mathbf{h}}+\mathbf{D}\frac{\partial \boldsymbol{\mathbf{{\mathbf{h}}}}}{\partial {\omega}}\right).
\end{equation}
%---------------------
Note that $\xi$ is regarded as constant when computing derivatives with respect to the LoS channel angles. 
From the definition of $\mathbf{D}$ and ${\mathbf{h}}$ in~\eqref{y}, we first need to calculate the partial derivative of $d_{n,m}^{-1}$ and $(\mathbf{h}_m)_{[n]}$, $\forall n\in \setI_N, m\in \setI_{|\setM|}$, based on~\eqref{eq:r_nm} and~\eqref{Gh}, respectively. We have
%-------------------------
\begin{align}
\label{doh_h_doh_omega}
\frac{\partial (\mathbf{h}_{m})_{[n]}}{\partial {\omega}}= \frac{\partial \textit{exp} \Big(\frac{-\textsf{j}2\pi d_{{n,m}}} {\lambda}\Big)}{\partial {\omega}}
=\frac{-\textsf{j}2\pi}{\lambda}\frac{\partial d_{n,m}}{\partial \omega}(\mathbf{h}_{m})_{[n]},
\end{align}
%---------------------
%-------------------------
\begin{equation}
\label{doh_D_doh_omega}
\frac{\partial d_{n,m}^{-1}}{\partial {\omega}}= -d_{n,m}^{-2}\frac{\partial d_{n,m}}{\partial \omega},
\end{equation}
%---------------------
where, it is noted from~\eqref{eq:r_nm} that $d_{n,m}$ depends on $(d_{1,1},\omega_{1,1},\gamma)$, in which $d_{1,1}$ is itself a function of $(\omega_{1,1},\omega_{1,M})$ as seen from~\eqref{eq:r11}. Consequently, the partial derivatives $\frac{\partial d_{n,m}}{\partial \omega_{1,1}}$ and $\frac{\partial d_{n,m}}{\partial \omega_{1,M}}$ are given in~\eqref{doh_d_doh_omega11} and~\eqref{doh_d_doh_omega1M}, respectively, at the top of this page, along with
% where noting the fact that from~\eqref{eq:r_nm}, $d_{n,m}$ depends on $(d_{1,1},\omega_{1,1},\gamma)$, where $d_{1,1}$ is a function of $(\omega_{1,1},\omega_{1,M})$ as seen from~\eqref{eq:r11}, the 
% $\frac{\partial d_{n,m}}{\partial \omega_{1,1}}$ and $\frac{\partial d_{n,m}}{\partial \omega_{1,M}}$ are given in~\eqref{doh_d_doh_omega11} and~\eqref{doh_d_doh_omega1M} at the top of this page, and
%...................................
\begin{align}
\addtocounter{equation}{+2} 
    \frac{\partial d_{n,m}}{\partial \gamma} &= \frac{l_{\text{UE},n}}{d_{n,m}} \bigg(l_{\text{BS},m}\sin\gamma-d_{1,1}\cos(\omega_{1,1}+\gamma)\bigg).
\end{align}
%-----------
Substituting~\eqref{doh_h_doh_omega} and~\eqref{doh_D_doh_omega} into~\eqref{doh_mu_doh_omega1}, we obtain
%-------------------------
\begin{align}
\label{doh_mu_doh_omega2}
\frac{\partial \boldsymbol{\mu}}{\partial {\omega}}&= %-\xi\Big(\tilde{\mathbf{D}}_{\omega}\mathbf{D}^2{\mathbf{h}}+\frac{\textsf{j}2\pi}{\lambda}\tilde{\mathbf{D}}_{\omega}\mathbf{D}{\mathbf{h}}\Big)\nonumber \\&=
-\xi\tilde{\mathbf{D}}_{\omega}\Big(\mathbf{D}+\frac{\textsf{j}2\pi}{\lambda}\mathbf{I}_{N|\setM|}\Big)\mathbf{D}{\mathbf{h}},
\end{align}
%---------------------
where $\tilde{\mathbf{D}}_{\omega}\in \mathbb{R}^{N|\setM|\times N|\setM|}$ is a diagonal matrix with $\frac{\partial d_{n,m}}{\partial {\omega}}$ on its diagonal.
Finally, by defining $\mathbf{G}_{\omega}\triangleq-\tilde{\mathbf{D}}_{\omega}\Big(\mathbf{D}+\frac{\textsf{j}2\pi}{\lambda}\mathbf{I}_{N|\setM|}\Big)$ and
using the fact that
 $\frac{\partial \boldsymbol{\mu}^\herm}{\partial \boldsymbol{\varepsilon}_{[i]}}=(\frac{\partial \boldsymbol{\mu}}{\partial \boldsymbol{\varepsilon}_{[i]}})^\herm$, the elements of FIM can be expressed as
% For this purpose, let us define $\boldsymbol{g}_m^{(\varepsilon)}$ corresponding to the parameter vector $\varepsilon$ as
% %-----------------------
% \begin{align} 
% \label{gm}
% \boldsymbol{g}_m^{(\varepsilon)}  \triangleq \bigg[(\frac{-1}{d_{1,m}}-\frac{2\textsf{j}\pi}{\lambda})\frac{\partial d_{1,m}}{\partial \varepsilon},\cdots,(\frac{-1}{d_{N,m}}-\frac{2\textsf{j}\pi}{\lambda})\frac{\partial d_{N,m}}{\partial \varepsilon}\bigg]^{\tran}.
% \end{align}
% %------------------------
% We also define $\mathbf{G}^{(\varepsilon)}\triangleq \diag\bigg(\boldsymbol{g}_{\setM(1)}^{(\varepsilon)},\cdots,\boldsymbol{g}_{\setM(|\setM|)}^{(\varepsilon)}\bigg)$. 
%------------------
\begin{equation}
\label{F2ij_1}    \mathbf{F}_{[i,j]} =  \frac{2|\xi|^2}{\tilde{\sigma}^2} \mathbf{1}^\tran\mathbf{G}_{\boldsymbol{\varepsilon}_{[i]}}^\herm \mathbf{D}^2 \mathbf{G}_{\boldsymbol{\varepsilon}_{[j]}} \mathbf{1}  ,\quad\forall (i,j)\in \setI_3.
\end{equation}
%----------------------
% \begin{align}
% \label{Fij_2}
%     &\F_{[i,j]} =\F_{[j,i]}=   \frac{2}{\tilde\sigma^2}\boldsymbol{1}^\tran \D^2 \nonumber\\&\bigg(\Re[{\xi}]\Re[\mathbf{G}^{(\boldsymbol{\varepsilon}_{[i]})}]+ \Im[{\xi}]\Im[\mathbf{G}^{(\boldsymbol{\varepsilon}_{[i]})}]\bigg)\mathbf{1}
%     \nonumber\\ &\forall i\in \setI_3,\: j=4.
% \end{align}
%-----------------------
%----------------------
% \begin{align}
% \label{Fij_3}
%     &\F_{[i,j]} =\F_{[j,i]}=   \frac{2}{\tilde\sigma^2}\boldsymbol{1}^\tran \D^2 \nonumber\\&\bigg(\Re[{\xi}]\Im[\mathbf{G}^{(\boldsymbol{\varepsilon}_{[i]})}]+ \Im[{\xi}]\Re[\mathbf{G}^{(\boldsymbol{\varepsilon}_{[i]})}]\bigg)\mathbf{1}
%     \nonumber\\ &\forall i\in \setI_3,\: j=5.
% \end{align}
%-----------------------
% \begin{align}
% \label{Fij_4}
%     \F_{[4,5]} = \F_{[5,4]}= 0.
% \end{align}
%------------------------
\begin{align}
\label{Fij_4}
    \F_{[i,i]} =\frac{2 \boldsymbol{1}^\tran \mathbf{D}^2\boldsymbol{1}}{\tilde\sigma^2}, \quad i=4,5.
\end{align}
%******************************
\begin{figure*}[t]
%\hrule
\begin{align}
    \F_{[i,j]} =\F_{[j,i]} &=   \frac{2}{\tilde\sigma^2}\boldsymbol{1}^\tran \D^2\bigg(\Re[{\xi}]\Re[\mathbf{G}_{\boldsymbol{\varepsilon}_{[i]}}]+ \Im[{\xi}]\Im[\mathbf{G}_{\boldsymbol{\varepsilon}_{[i]}}]\bigg)\mathbf{1},\quad \forall i\in \setI_3,\: j=4, \label{Fij_3} \\
    \F_{[i,j]} =\F_{[j,i]} &=   \frac{2}{\tilde\sigma^2}\boldsymbol{1}^\tran \D^2 \bigg(\Re[{\xi}]\Im[\mathbf{G}_{\boldsymbol{\varepsilon}_{[i]}}]+ \Im[{\xi}]\Re[\mathbf{G}_{\boldsymbol{\varepsilon}_{[i]}}]\bigg)\mathbf{1},\quad \forall i\in \setI_3,\: j=5. \label{Fij_4}
\end{align}
\hrule
\end{figure*}
%*************************
Also, $\F_{[i,j]}$, $\forall i\in\setI_3, j=4,5$ are given 
in~\eqref{Fij_3} and~\eqref{Fij_4} on the top of the next page. Moreover we have
$\F_{[4,5]} = \F_{[5,4]}= 0$.
To provide the final CRLB of the LoS channel angles after the quantized feedback, the impact of the quantization granularity must also be taken into account. 
Let $\boldsymbol{\omega}_q=Q_q(\boldsymbol{\omega})$ denote the quantized version of the angular vector $\boldsymbol{\omega}$. %Assuming perfect quantization, we have $\boldsymbol{\omega}=Q_q^{-1}(\boldsymbol{\omega}_q)$, leading to the CRLB after the feedback phase as $\mathbf{CRB}=\mathbf{F}^{-1}$.
%However, i
In practice, finite quantization introduces error, such that $Q_q^{(-1)}(\boldsymbol{\omega}_q)=\boldsymbol{\omega}+\mathbf{e}$, where $\mathbf{e}$ represents the quantization error. Assuming that $Q_q(.)$ is a uniform quantizer with step size $\Delta_i$ for $\boldsymbol{\omega}_{[i]}$, the $i^\text{th}$ element of $\mathbf{e}$ follows a uniform distribution, $\mathcal{U}(-\frac{\Delta_i}{2},\frac{\Delta_i}{2})$. Thus, the CRLB of each quantized reference angle is given by
% To incorporate the impact of angle quantization for the feedback phase, we assume that the quantizer is uniform with step size $\Delta$, and the quantization error associated with each reference angle is denoted by $e_q$. Thus, for each angle $\omega \in \{\hat{\omega}_{1,1}
% ,\hat{\omega}_{1,M},\hat{\gamma}\}$ the received angle at the BS is given by
% \begin{align}
% \label{omega_e}
% Q(\omega)=\omega+e_q.
% \end{align}
% The quantization error follows the uniform distribution as $e_q\sim \mathcal{U}(-\frac{\Delta}{2},\frac{\Delta}{2})$, and the final FIM is given by
% \begin{align}
% \label{FIM}
% \mathbf{FIM}=\F+
% \diag\left(\frac{12}{\Delta_1^2},\frac{12}{\Delta_2^2},\frac{12}{\Delta_3^2},0,0\right)
% \end{align}
%where $\sigma_Q^2=\frac{\Delta^2}{12}$. 
\begin{align}
    \label{CRB}    &\mathbf{C}_{[i,i]}\leq(\mathbf{F}^{-1})_{[i,i]}+\frac{\Delta_i^2}{12}; \quad i\in \setI_3  
\end{align}
%\red{The analysis of the CRLB parameters for differnt geometric parameters is analyzed in Section....} 
In Section~\ref{sec:num} we employ these CRLB expressions to analyze the impact of pilot count, array size, and relative BS-UE orientation on estimation accuracy.

\section{Two-Stage Transmission Design Based on the Near-field LoS CSI and OTA Iterations}
\label{sec:Precoder_Design}
In this section, we discuss transmission design, focusing on precoding design at the BS using multiuser NF LoS CSI obtained through Algorithm~\ref{alg1}.
After acquiring LoS CSI for all $K$ UEs, the BS can design the precoding vectors $\{\{\mathbf{m}_{k,s}\}_{s=1}^{S_k}\}_{k=1}^K$ using the collected LoS channels $\{\hat{\mathbf{H}}_k^\text{LoS}\}_{k=1}^K$. However, relying solely on the estimated LoS channel may have several shortcomings. Noisy reference angle estimates at the UE or limited quantization levels can introduce significant errors in the reconstructed channel. Possible means for mitigation include increasing grid granularity, increasing the number of pilots, or refining quantization. However, these methods increase complexity and pilot overhead.
While NLoS components are often negligible at mmWave and THz due to high path loss, neglecting them at lower frequencies degrades channel reconstruction. Accurate NLoS estimation also incurs substantial pilot and feedback overhead.
%While NLoS components in mmWave and THz bands are often negligible due to high path loss, ignoring them in lower frequency bands leads to inaccurate channel reconstruction. Furthermore, accurately estimating the NLoS channel significantly increases the feedback and pilot overhead.
% %***********************
% \begin{figure}[t!]
%       \centering %\includegraphics[trim=0.30cm 0.18cm 0.22cm 0.22cm, width=\columnwidth]
%       \includegraphics[ width=\columnwidth]
%      {Figs/Procedure.PNG}
%     \caption{Signaling flow of two-stage TX-RX beamforming design.}
%    % \vspace{-0.5cm}
%     \label{fig:system_model2}
% \end{figure}
% %************************

\begin{figure}[t!]
      \centering %\includegraphics[trim=0.30cm 0.18cm 0.22cm 0.22cm, width=\columnwidth]
      \includegraphics[ width=\columnwidth]
     {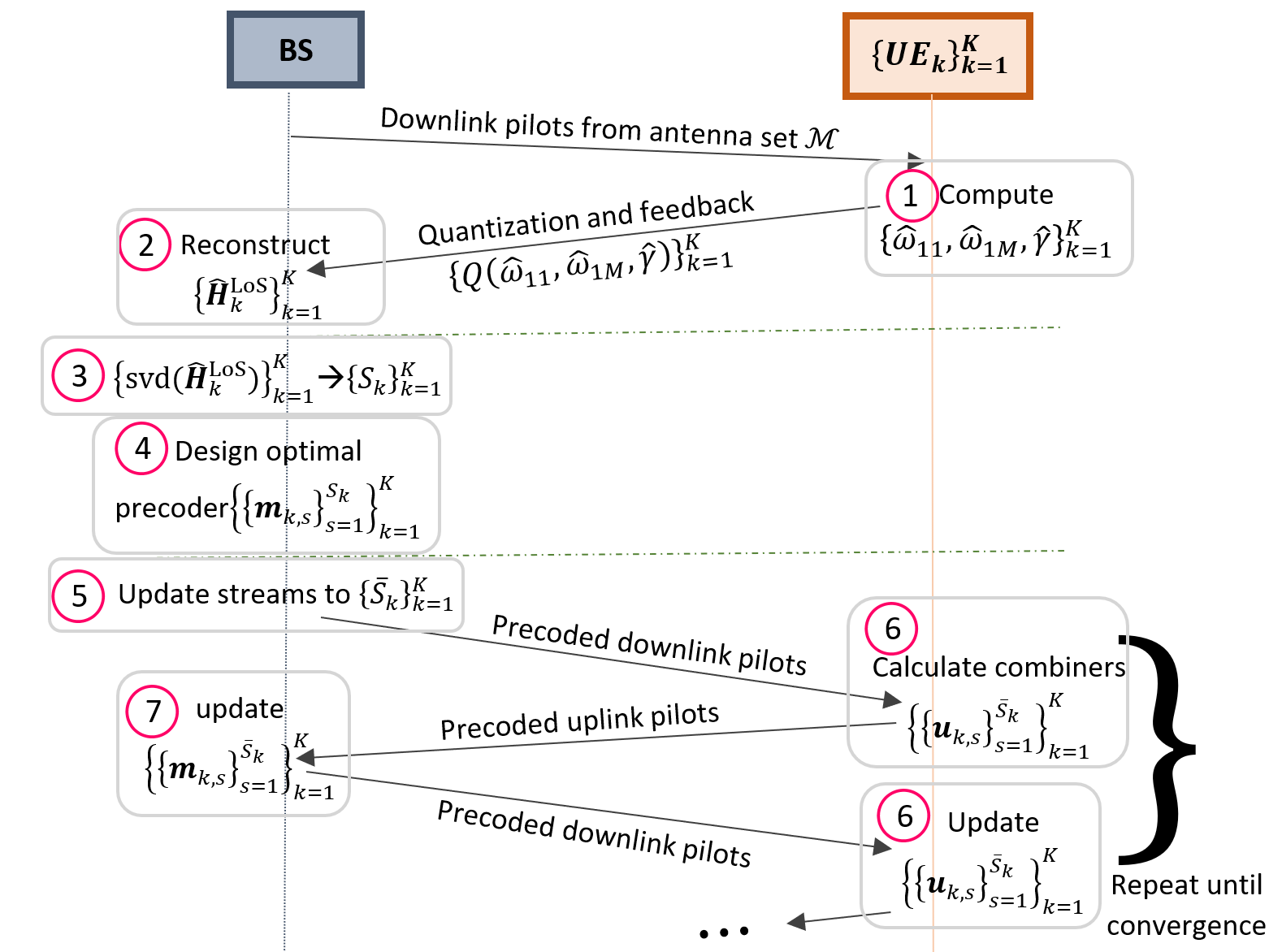}
    \caption{Proposed two-stage transmission design signaling.}
   % \vspace{-0.5cm}
    \label{fig:system_model2}
\end{figure}

To address these challenges, we propose a two-stage transmission design depicted in Figure~\ref{fig:system_model2}. In the first stage, the BS designs the precoding vectors $\{\{\mathbf{m}_{k,s}\}_{s=1}^{S_k}\}_{k=1}^K$ based on the estimated LoS channels $\{\hat{\mathbf{H}}_k^\text{LoS}\}_{k=1}^K$, serving as a coarse precoding refined in the second stage via OTA iterations.
Before both stages, a stream allocation process is introduced to accelerate precoding convergence and minimize OTA pilot overhead. In the first phase, the BS determines the stream count~$S_k$ for each UE $k$ based on the rank of $\hat{\mathbf{H}}_k^\text{LoS}$. Before the OTA phase, $S_k$ is updated to $\bar{S}_k \leq S_k$ by eliminating near-zero-rate streams, computed using coarse precoding and LoS CSI. %****************
 The OTA stage begins with the downlink transmission of pilot sequences precoded by the coarse precoding vectors $\{\{\mathbf{m}_{k,s}\}_{s=1}^{\bar{S}_k}\}_{k=1}^K$.  Each
UE $k$ computes its combining vectors $\{\mathbf{u}_{s,k}\}_{s=1}^{\bar{S}_k}$ from the received downlink signal. The UEs then use these vectors to precode the pilot sequences and transmit them in the uplink. This process allows the BS to refine the precoding based on the received uplink signal and proceed with the subsequent OTA downlink transmission similarly to~\cite{Antti_BiT}.
Note that both the forward and backward transmissions are performed over the actual physical channel, which enables refinement of the precoding and combining through OTA training iterations that inherently account for both LoS and NLoS components.
%Note that both the forward and backward transmissions are carried out through the actual physical channel, allowing the precoding and combining to be refined via OTA training iterations. These iterations naturally account for both LoS and NLoS components.
%-------------------------------------------------------------------------------------
%Another version:
% Finally, the second stage is initiated by transmitting precoded pilots in the downlink using $\{\{\mathbf{m}_{k,s}\}_{s=1}^{\bar{S}_k}\}_{k=1}^K$, where $\bar{S}k$ is the updated number of streams for the $k^{\text{th}}$ user after the second stream filtering. Each user $k$  then obtains its combiner vectors $\{\mathbf{u}_{s,k}\}_{s=1}^{\bar{S}_k}$ by extracting the equivalent channel from the received signal. In the uplink, each user transmits backward by precoding the pilots with $\{\mathbf{u}_{s,k}\}_{s=1}^{\bar{S}_k}$, allowing the BS to update the precoders by extracting the equivalent multiuser channel. Note that both forward and backward transmissions contain the true channel information, enabling refinement of the precoders/combiners through forward/backward OTA training iterations that account for both LoS and NLoS components. 
%--------------------------------------------------------------------------------
%In the following, we provide a detailed description of each stage. 
\subsection{Coarse Initial Precoding Design Using NF LoS CSI}
%*************************
%\subsubsection{Stream Allocation}
%*************************
The coarse precoding design starts with stream allocation, where the BS allocates $S_k$ streams to each UE~$k$ by performing the singular value decomposition (SVD) of $\hat{\H}_k \triangleq \hat{\H}^\text{LoS}_k$, given by $\hat{\mathbf{H}}_k={\mathbf{W}}_k\boldsymbol{\Delta}_k{\mathbf{V}}_k^\herm$, where $\boldsymbol{\Delta}_k=\diag(\delta_{k,1},\cdots, \delta_{k,N})$ contains the singular values of $\hat{\mathbf{H}}_k$, and $\mathbf{W}_k$ and ${\mathbf{V}}_k$ are the left and right singular matrices, respectively.
% The coarse precoding design procedure starts with the stream allocation process. In this step, the BS allocates $S_k$ streams to each UE $k$ by performing the singular value decomposition (SVD) of $\hat{\H}_k \triangleq \hat{\H}^\text{LoS}_k$, which is given as $\hat{\mathbf{H}}_k={\mathbf{W}}_k\boldsymbol{\Delta}_k{\mathbf{V}}_k^\herm$, where $\boldsymbol{\Delta}_k=\diag(\delta_{k,1},\cdots, \delta_{k,N})$ contains the singular values $\hat{\mathbf{H}}_k$, and $\mathbf{W}_k$ and ${\mathbf{V}}_k$ are the left and right singular matrices, respectively. 
The BS then selects the first $S_k$ singular values based on a predefined criterion.
For instance, the stream corresponding to the singular value $\delta_{k,n}$ is selected if
$\delta_{k,n}\geq \delta_{k}^\star/\alpha$, where $\delta_{k}^\star = \mathop{\text{max}}\limits_{n \in \setI_N} \delta_{k,n}$ represents the largest singular value, and $\alpha$ is a predefined threshold. Since the stream allocation is based on estimated LoS channels, the threshold 
$\alpha$ should be chosen conservatively to ensure that most of the eigenmodes are retained. This approach helps avoid discarding useful eigenmodes that may still contribute valuable information, despite the estimation error. For example, $\alpha = [10,100,1000]$ values may be used depending on the distribution of singular values. 
%*************************
%\subsubsection{Optimal Precoder Design}
%*************************
%Now, we describe the optimization problem for obtaining the precoding vectors based on the LoS CSI $\{\hat{\mathbf{H}}_k\}_{k=1}^K$ collected at the BS. 
We now formulate the optimization problem for precoding design at the BS based on the LoS CSI $\{\hat{\mathbf{H}}_k\}_{k=1}^K$. In general, the precoding/combining vectors can be obtained by solving %the following problem
%--------------------------
 \begin{equation}
\label{eq:Optimization_problem1}
\begin{aligned}
 \min_{\u_{k,s}, \m_{k,s}} &\quad f_\text{o}(\hat{\Gamma}_{k,s}) \\ \text{s. t.}&  \quad \sum_{k=1}^K\sum_{s=1}^{S_k} \|\m_{k,s}\|^2 \leq P_\text{BS},
\end{aligned}
\end{equation}
%---------------------------
where $f_\text{o}(\hat{\Gamma}_{k,s})$ is the objective function for the linear beamformer design, $\hat{\Gamma}_{k,s}$ is obtained by replacing ${\mathbf{H}_k}$ with $\hat{\mathbf{H}}_k$ in \eqref{eq:SINR_sk}, and $P_\text{BS}$ is the maximum BS power. Typically,~\eqref{eq:Optimization_problem1} is not a jointly convex function of the optimization variables $\u_{k,s}$ and $\m_{k,s}$~\cite{Jarkko,Nariman, Ganesh} and hence is solved approximately through an iterative process.  For a fixed $\{\{\mathbf{m}_{k,s}\}_{s=1}^{S_k}\}_{k=1}^K$, the combining vectors for all $(k,s)$ are obtained as~\cite{Jarkko}
%****************
\begin{equation}
\label{eq:u_ks}
    \begin{aligned}    \u_{k,s}=\bigg(\sum_{\bar k=1}^K \sum_{\bar s=1}^{S_{\bar k}}
 \hat{\H}_k \m_{\bar k,\bar s}\m_{\bar k, \bar s}^{\herm} \hat{\H}_k^{\herm} + \sigma_k^2 \I\bigg)^{-1}\hat{\H}_k \m_{k,s}.   
\end{aligned}
\end{equation}
%***************************
For typical objective functions, such as,
$f_0(\hat{\Gamma}_{k,s})=-\min_{k \in \setI_K} \sum_{s=1}^{S_k} \log(1+\hat{\Gamma}_{k,s})$ considered in this paper, and for a fixed set of combining vectors $\{\{\mathbf{u}_{k,s}\}_{s=1}^{S_k}\}_{k=1}^K$, the precoding vectors can be expressed as~\cite{Jarkko,Nariman}
%******************************************
%******************************
%***********************************************
%*************************
\begin{equation}
\label{eq:m_kl}
    \begin{aligned}  \m_{k,s}=\bigg(\sum_{\bar k=1}^{K} \sum_{\bar s=1}^{S_{\bar{k}}}
\nu_{\bar{k},\bar{s}}\hat{\mathbf{h}}^{\textit{eff}}_{\bar{k},\bar{s}} \left(\hat{\mathbf{h}}^{\textit{eff}}_{\bar{k},\bar{s}}\right)^{\herm}+ \eta \I\bigg)^{-1} \nu_{{k},{s}}\hat{\mathbf{h}}^{\textit{eff}}_{k,s},  
\end{aligned}
\end{equation}
%******************************************
where, $\{\{\nu_{k,s}\}_{s=1}^{S_k}\}_{k=1}^K$ are the dual variables that satisfy the Karush-Kuhn-Tucker (KKT) conditions of the convexified version of~\eqref{eq:Optimization_problem1}.
The term $\hat{\mathbf{h}}^{\textit{eff}}_{k,s}\triangleq\hat{\H}_k^{\herm}\u_{k,s}$ denotes the effective uplink channel for stream $s$ of UE $k$ 
%given a fixed set of combining vectors $\u_{k,s}$, 
computed based on the estimated LoS CSI, and $\eta$ is obtained via a bisection method to enforce the power constraint. See Appendix~\ref{app:A3} for further details on solving~\eqref{eq:Optimization_problem1}.
In summary, coarse precoding design at the BS using estimated LoS CSI follows an iterative precoding-combining update process.  The process begins by generating an initial feasible precoder set, $\big\{\{\mathbf{m}_{k,s}\}_{s=1}^{S_k}\big\}_{k=1}^K$, that satisfies the power constraint. The initial precoding associated with the UE $k$ can be selected as the (scaled) right singular vectors of $\hat{\mathbf{H}}_k$, i.e., $c_1\big({\mathbf{V}}_k^\herm\big)_{[:,1:S_k]}$, where the scaling factor
$c_1$ ensures compliance with the power constraint and is given by
\begin{equation}
\label{eq:c1}
    \begin{aligned}  c_1=\frac{P_\text{BS}}{\sum_{k=1}^K\|({\mathbf{V}}_k^\herm\big)_{[:,1:S_k]}\|^2_\text{F}}
    .
\end{aligned}
\end{equation}
The combining vectors are then computed using \eqref{eq:u_ks}. Next, the precoding vectors are updated via \eqref{eq:m_kl}, assuming fixed combining. This iterative precoding-combining update process continues until convergence. The final precoding vectors obtained at the end of this phase serve as the initial precoding for the OTA procedure.
%***********************
\subsection{Precoding Refinement via Bi-directional OTA Training}
% In general, precoder/combiner design via bidirectional OTA training is performed without explicit channel estimation. This procedure typically begins with a set of random precoders (or combiners) and is refined through iterative downlink and uplink pilot transmissions, which inherently include the true channel. However, we utilize a coarse precoder designed at the BS based on the LoS CSI as the initial precoder for the OTA phase. This approach reduces pilot overhead and the number of iterations required during the OTA phase, while also accounting for the impact of the NLoS channel on transmission design. In this section, we describe this procedure in detail.
%\subsubsection{Stream Filtering}
To refine the precoding obtained from estimated LoS CSI in the previous phase, we employ iterative bidirectional OTA training to capture the impact of NLoS components.
To further reduce OTA iterations and pilot overhead, 
The BS allocates streams by evaluating stream-specific rates using the estimated LoS CSI and the precoding/combining from the previous stage.
%the BS performs a stream allocation by evaluating stream-specific rates using the estimated LoS CSI along with the precoding and combining from the previous stage.
For each UE $k$ the rate of the $s^\text{th}$ stream is estimated as $\hat{r}_{k,s}=\log_2(1+\hat{\Gamma}_{k,s}), \forall s\in \setI_{S_k}$, where $\hat{\Gamma}_{k,s}$ is computed using $\{\{\mathbf{m}_{k,s}\}_{s=1}^{S_k}\}_{k=1}^K$ and $\mathbf{u}_{k,s}$ obtained from the coarse precoding design procedure. Streams with near-zero rates are then removed from $\setI_{S_k}$, adjusting the allocated streams to $\bar{S}_k\leq S_k$. To meet the power constraint, the precoding vectors $\{\{\mathbf{m}_{k,s}\}_{s=1}^{\bar{S_k}}\}_{k=1}^K$ may require scaling with factor $c_2$, where
    \begin{align}\label{eq:c2}
  c_2=\frac{P_\text{BS}}{\sum_{k=1}^K\sum_{k=1}^{\bar{S}_k}\|{\mathbf{m}}_{s,k}\|^2}.
\end{align}
%******************************
% \begin{figure*}[t]
%     \centering
%     % First subfigure
%     \begin{subfigure}[b]{0.3\textwidth} % Set each subfigure width to 0.3\textwidth
%         \centering
%         \includegraphics[width=\linewidth]{Fig2/CRLB_MSE_omega.eps} % Adjust width to fill the subfigure
%         \caption{CRLB, MSE, and UE achievable rate computed based on the LoS CSI vs. number of pilots for different quantization levels.}
%         \label{fig:CRB_MSE_omega}
%     \end{subfigure}
%     \hspace{0.01\textwidth} % Adjust horizontal spacing between subfigures
%     % Second subfigure
%     \begin{subfigure}[b]{0.3\textwidth}
%         \centering
%         \includegraphics[width=\linewidth]{Fig2/CRLB_MSE_gamma.eps}
%         \caption{Impact of the UE antenna array length on the CRLB of LoS angles for different numbers of pilots.}
%         \label{fig:CRB_MSE_gamma}
%     \end{subfigure}
%     \hspace{0.01\textwidth} % Adjust horizontal spacing between subfigures
%     % Third subfigure
%     \begin{subfigure}[b]{0.3\textwidth}
%         \centering
%         \includegraphics[width=\linewidth]{Fig2/Rate.eps}
%         \caption{The objective function of the LoS angle estimator as a function of $(\omega_{1,M}, \gamma)$, with $\omega_{1,1} = \hat{\omega}_{1,1}$, for a single UE realization.}
%        % and a grid step size of $0.1^\circ$.
%         \label{fig:Rate}
%     \end{subfigure}
%      \hspace{0.01\textwidth} % 
%     \caption{Evaluating LoS angles estimation accuracy via CRLB, MSE, rate, and the objective function of the estimator.}
%     \label{fig:CRB_pilot}
% \end{figure*}

\begin{algorithm}
%\begin{algorithmic}
%\color{gray}
  \textbf{Data:} $\{\hat{\mathbf{H}}_k\}_{k=1}^K$, $\{\sigma^2_k\}_{k=1}^K$, $P_\text{BS}$,  $\alpha$.
  %$\{\{\tilde{\mathbf{p}}_{k,s}\}_{s=1}^{{S_k}}\}_{k=1}^K$, 
 % \hspace{8mm}$\{l_{\text{TX},m}\}_{j=1}^M$ and $\{l_{\text{RX},i}\}_{i=1}^N$, respectively.
 
 \vspace{1mm}
 \textbf{Stage 1: Coarse precoding design at the BS}:
%****************************
\begin{itemize}[leftmargin=12mm]
\item[\small\texttt{(S.0)}] Decompose $\hat{\mathbf{H}}_k$ via SVD as $\mathbf{W}_k \boldsymbol{\Delta}_k \mathbf{V}_k^\herm$, and define $S_k$ as the number of singular values greater than $\delta_k^\star=\alpha^{-1} (\boldsymbol{\Delta}_k)_{[1,1]}$, $\forall k$.
\item[\small\texttt{(S.1)}]Initialize precoding vectors as $\{\{\mathbf{m}_{k,s}\}_{s=1}^{S_k}\}_{k=1}^K=\{\{c_1\mathbf({V}_k^\herm)_{[:,s]}\}_{s=1}^{S_k}\}_{k=1}^K$, where $c_1$ is given in~\eqref{eq:c1}.
\item[\textbf{While}]
\textit{convergence criterion is not met} \textbf{do}
\item[\small\texttt{(S.2)}] Calculate $\mathbf{u}_{k,s}$ from \eqref{eq:u_ks} using $\hat{\mathbf{H}}_k$,  $\forall (k,s)$. 
\item[\small\texttt{(S.3)}] Update $\mathbf{m}_{k,s}$ from~\eqref{eq:m_kl} or via CVX solution, $\forall (k,s)$.
%(see Appendix~\ref{app:A3}).
 \item[\textbf{End}]
\end{itemize}
%*******************************
%*****************************
\textbf{Stage 2: Precoding refinement via OTA iterations}:
\begin{itemize}[leftmargin=12mm] 
\item[\small\texttt{(S.4)}] Calculate $\hat{r}_{k,s}=\log_2(1+\hat{\Gamma}_{k,s})$ via~\eqref{eq:SINR_sk} using the latest $\mathbf{u}_{k,s}$ and $\{\{\mathbf{m}_{k,s}\}_{s=1}^{S_k}\}_{k=1}^K$,  based on $\hat{\mathbf{H}}_k$, $\forall (k,s)$.
\item[\small\texttt{(S.5)}] For each UE $k$, discard streams with $\hat{r}_{k,s} \simeq 0$, and update the number of streams as $S_k \leftarrow \bar{S}_k$.
% \item Scale $\{\{{\mathbf{m}}_{k,s}\}_{s=1}^{{\bar{S}_k}}\}_{k=1}^K$ by $c_2$ to meet the power constraint.
 \item[\small\texttt{(S.6)}] set ${\mathbf{m}}_{k,s}\leftarrow c_2 
{\mathbf{m}}_{k,s},~\forall (k,s)$, using $c_2$ from~\eqref{eq:c2}.
\item[\textbf{While}]
\textit{convergence unmet} \textbf{do} (standard OTA process~\cite{Biksh})
%*****************************
\item[\small\texttt{(S.7)}]
\textbf{BS:} Precode
pilot sequences by
$\{\{{\mathbf{m}}_{k,s}\}_{s=1}^{{\bar{S}_k}}\}_{k=1}^K$  and transmit in the downlink.
%as \eqref{eq:Yk}.
%*****************************
\item[\small\texttt{(S.8)}]
\textbf{UEs:} Calculate $\{{\mathbf{u}}_{k,s}\}_{s=1}^{{\bar{S}_k}}$ using $\mathbf{Y}_k^\text{fwd}$, $\forall k$.
\item[\small\texttt{(S.9)}]
\textbf{UEs:} Precode pilot sequences by
$\{{\mathbf{u}}_{k,s}\}_{s=1}^{{\bar{S}_k}}$, $\forall k$ and transmit in the uplink. 
%as \eqref{eq:Y_bwd}.
%*****************************
% \item[\small\texttt{(S.10)}] \textbf{BS:}
% Extract $\mathbf{h}_{k,s}^\textit{eff},~\forall(k,s)$ from \eqref{eq:eqch} using $\mathbf{Y}^\text{bwd}$. 
%*****************************
\item[\small\texttt{(S.10)}]\textbf{BS:}
Update $\mathbf{m}_{k,s}  \forall(k,s)$  using $\mathbf{Y}^\text{bwd}$.
%*****************************
\item[\textbf{End}]
 \end{itemize}
 %\end{algorithmic}
\caption{Proposed two-stage transmission design} \label{alg_5}
\end{algorithm}
%_____________________________________
%After updating the stream counts and initializing the refinement process with the coarse precoding from the previous phase, the OTA training phase iterates, following the standard bi-directional training procedure. 
After updating the stream counts and initializing the refinement process with the coarse precoding, the OTA phase proceeds iteratively via standard bi-directional training. 
%described in~\cite{Biksh, Antti_BiT, BDT,Praneeth}. 
Specifically, the BS precodes the pilot sequences using $\{\{c_2\mathbf{m}_{k,s}\}_{s=1}^{\bar{S_k}}\}_{k=1}^K$ and transmits them in the downlink. The subsequent steps proceed similarly to~\cite{Biksh, Antti_BiT, BDT, Praneeth}, where UE~$k$ receives the forward signal as~\cite{Biksh}
%*******************************************
    \begin{align}\label{eq:Yk}
    \Y_k^{\text{fwd}}= \H_k\sum_{\bar k=1}^K \sum_{\bar s=1}^{\bar{S}_k}  \m_{\Bar{k},\Bar{s}}\tilde{\mathbf{p}}_{\Bar{k},\Bar{s}}^{\herm} + \Z_k \in \mathbb{C}^{N \times \tilde{\tau}},\quad \forall k,
\end{align}
where $\Z_k$ is the noise at the UE $k$ with the entries each distributed as $\mathcal{CN} (0,\sigma_{k}^2)$, and $\tilde{\mathbf{p}}_{k,s}\in \mathbb{C}^{\tilde{\tau}}$ is the pilot vector. The UE $k$ estimate the combining vectors~\eqref{eq:u_ks} from $\mathbf{Y}_k^\text{fwd}$, which contains the true channel $\mathbf{H}_k = \mathbf{H}_k^\text{LoS} + \mathbf{H}_k^\text{NLoS}$, i.e.,~\cite{Biksh}
%The combining vectors~\eqref{eq:u_ks} are estimated by the UE using $\mathbf{Y}_k^\text{fwd}$, which contains the true channel $\mathbf{H}_k=\mathbf{H}_k^\text{LoS}+\mathbf{H}_k^\text{NLoS}$, i.e.,~\cite{Biksh}
%*******************************************
\begin{equation}
\label{eq:u_fwd}
    \begin{aligned}       \u_{k,s}\approx\Big(\Y_k^{\text{fwd}}(\Y_k^{\text{fwd}})^\herm\Big)^{-1} \Y_k^{\text{fwd}}\tilde{\mathbf{p}}_{k,s} ,\:\: \forall (k,s).
\end{aligned}
\end{equation}
%*******************************************
The UEs perform backward transmission by precoding the pilots using~\eqref{eq:u_fwd}, after which the BS receives $\Y^{\text{bwd}}$ as~\cite{Biksh}
 %*******************************************
\begin{equation}
\label{eq:Y_bwd}
    \begin{aligned}    \Y^{\text{bwd}}=\psi\sum_{k=1}^K \H_k^{\herm}  \sum_{\bar s=1}^{\bar{S}_k} \u_{k,s}\tilde{\mathbf{p}}_{k,s}^{\herm} + {\mathbf{Z}}_{\text{BS}} \in \mathbb{C}^{M \times \tilde{\tau}},
\end{aligned}
\end{equation}
%*******************************************
where $\Z_{\text{BS}}$ is the noise at the BS with element distributed as  $\mathcal{C} \mathcal{N} (0,\sigma_\text{BS}^2)$, and $\psi$ is a common scaling factor to meet the UEs power constraint. 
Assuming orthogonal pilots, the BS estimates the effective uplink channel as ${\h}^\textit{eff}_{k,s} \approx \frac{1}{\psi \tilde{\tau}} \Y^{\text{bwd}} \tilde{\mathbf{p}}_{k,s}^\text{H}$, and updates the precoding vectors by replacing $\hat{\h}^\textit{eff}_{k,s}$ with ${\h}^\textit{eff}_{k,s}$ in~\eqref{eq:m_kl}. See~\cite{Praneeth} for signaling details. Algorithm~\ref{alg_5} summarizes the proposed two-stage precoding
design procedure.
\section{Numerical Results and Discussion}
\label{sec:num}
In this section, we evaluate the performance of the proposed LoS CSI acquisition and transmission design algorithms through numerical examples. We consider the configuration illustrated in Figure~\ref{fig:system_model1}, where the BS is equipped with $M=64$ equally spaced antenna elements, spanning a total length of $l_{\text{BS},M}=2\text{m}$. The BS serves $K=8$ UEs, each equipped with $N$ antennas spaced at half-wavelength intervals, with a wavelength of $\lambda=3~\text{cm}$. The transceivers are located within a room with dimensions $15\text{m} \times 17\text{m} \times 8\text{m}$.
% %*****************************
% \begin{figure}[h]
%     \centering
%     \begin{subfigure}[b]{0.7\linewidth}
%         \centering
%         \includegraphics[width=\linewidth]{Fig2/Likelihood_2P.eps}
%         \caption{CRLB L-UE}
%     \end{subfigure}   
%     \begin{subfigure}[b]{0.7\linewidth}
%         \centering
%         \includegraphics[width=\linewidth]{Fig2/Likelihood_4P.eps}
%         \caption{MSE versus SNR for target 2.}
%     \end{subfigure}
%     \caption{CRLB Ltx.}
%     \label{fig:MSE_SNR}
% \end{figure}
% %*************************
The center of the BS array is assumed to be positioned at %$(-15\text{m}, 8.5\text{m}, 4 \text{m})$, 
$(0, 0, 1.5 \text{m})$, as seen in Figure~\ref{fig:system_model1}, while the UEs are randomly located on a horizontal plane at $z_0=1.5\text{m}$, i.e, $-8.5\text{m}\leq y_m^\text{BS}, y_n^{\text{UE}_k}\leq8.5\text{m}$, $x_m^\text{BS}=0$, and $0< x_n^{\text{UE}_k}\leq7.5\text{m}$, $\forall m \in \setI_M,~\forall n\in\setI_N, \forall k\in \setI_K$. To control the location of each UE within the plane $z=z_0$, we define $d_k$ as the distance between the centers of the BS and UE $k$. We also define $\beta_k$ as the angle between $d_k$ and the normal to the BS array. This angle, along with $d_k$ and $\gamma_k$, governs the relative location and rotation of the UE $k$ with respect to the BS. For example, $\gamma_k=0$ and $\beta_k=\pi/2$ correspond to a perfectly aligned LoS scenario, while setting $\beta_k \neq 0$ moves the UE array along the $y$-axis.

We generate random locations for each UE by sampling $d_k \in [4.5~\text{m}, 8.5~\text{m}]$, $\gamma_k \in [0, \pi]$, and $|\beta_k| \leq 80^\circ$, $\forall k$. To model the NLoS components, we use the geometry-based model described in Section IV of \cite{Indoor_mmw}, where the NLoS components consist of first-order and second-order reflections from the side walls and ceiling.
We assume that the BS power budget for the precoding design phase is $P_\text{BS}=-5~\text{dBm}$, while the BS power for the channel acquisition phase is $\rho=-3~\text{dBm}$. %specified for each experiment. 
The noise variance at UE $k$ and the BS is given by ${\sigma}^2_{k}= {\sigma}^2_{\text{BS}} = -85~\text{dBm}$, respectively, for both the CSI acquisition and precoding/combining design phases. Pilot sequences are generated using a DFT matrix, where the sequence length is set to $\tau=4$ for $|\setM|\leq4$ and $\tau=|\setM|$ for $|\setM|\geq4$, during the parameter estimation stage. In the OTA phase, the pilot length is defined as $\tilde{\tau}=M$. We define $\Delta\omega$ as the search grid step size used to construct the angular grid $\boldsymbol{\setA}$, which is set to $\Delta\omega=0.08^\circ$, unless otherwise specified.
%*****************************
\begin{figure}[h]
    \centering
    \begin{subfigure}[b]{1\linewidth}
        \centering        \includegraphics[width=\linewidth]{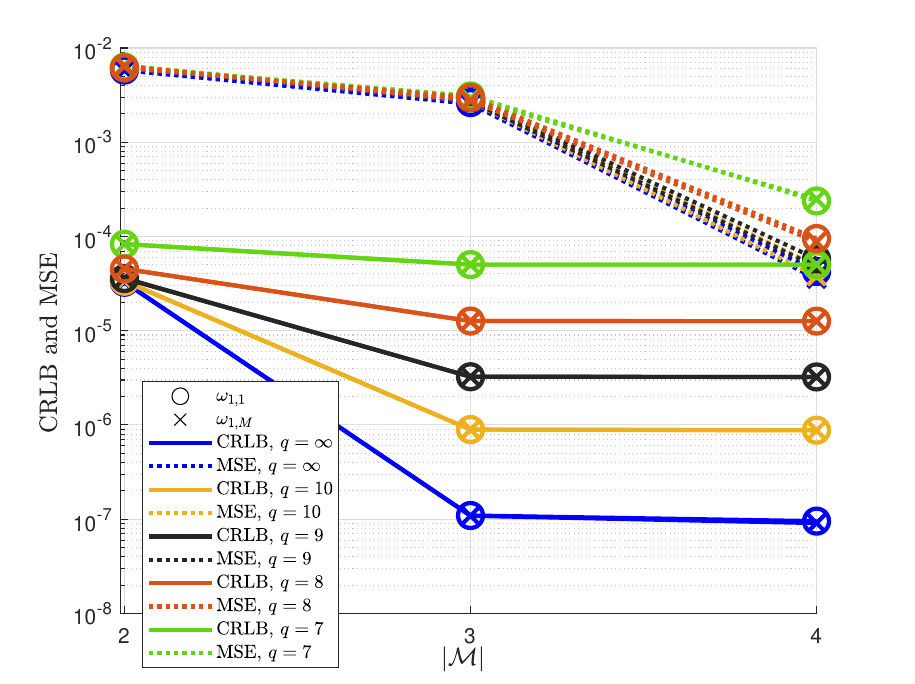}
        \caption{CRLB and MSE of $\omega_{1,1}$ and $\omega_{1,M}$ vs. number of pilot-transmitting BS antennas for various quantization bits of $\boldsymbol{\omega}$.}
        \label{fig:CRLB_omega111}
    \end{subfigure}    
    \begin{subfigure}[b]{1\linewidth}
        \centering        \includegraphics[width=\linewidth]{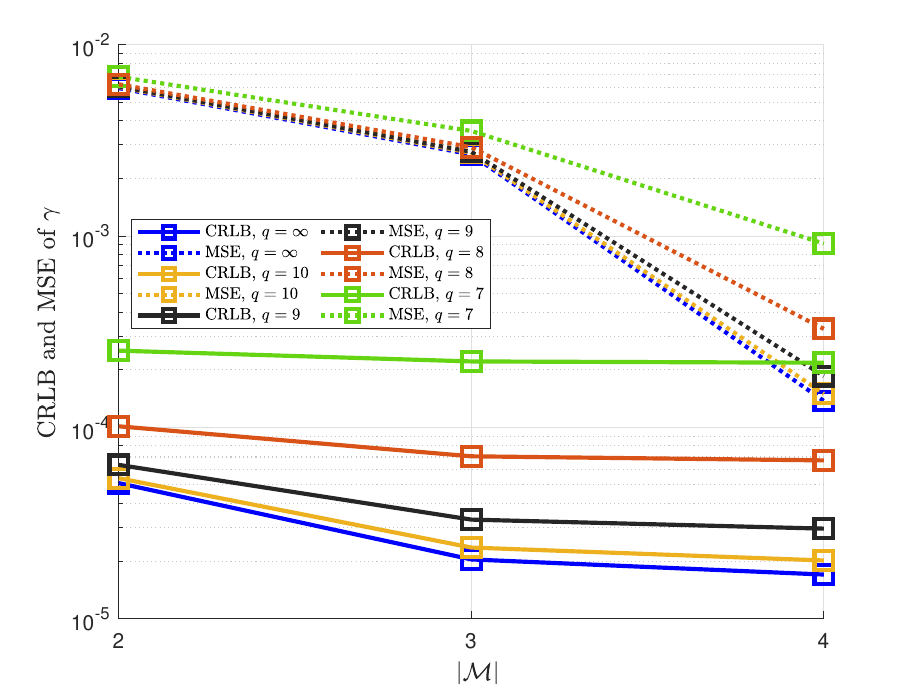}
        \caption{CRLB and MSE of  $\gamma$ vs. number of pilot-transmitting BS antennas $|\setM|$ for different quantization level.}
        \label{fig:CRLB_gamma}
    \end{subfigure}
    \caption{CRLB and MSE for LoS angles estimation}
        %\vspace{-0.5cm}
    \label{fig:MSE_CRLB}
\end{figure}
%***************************
%****************************
\begin{figure}[t!]
    \centering
    \includegraphics[width=1 \columnwidth]{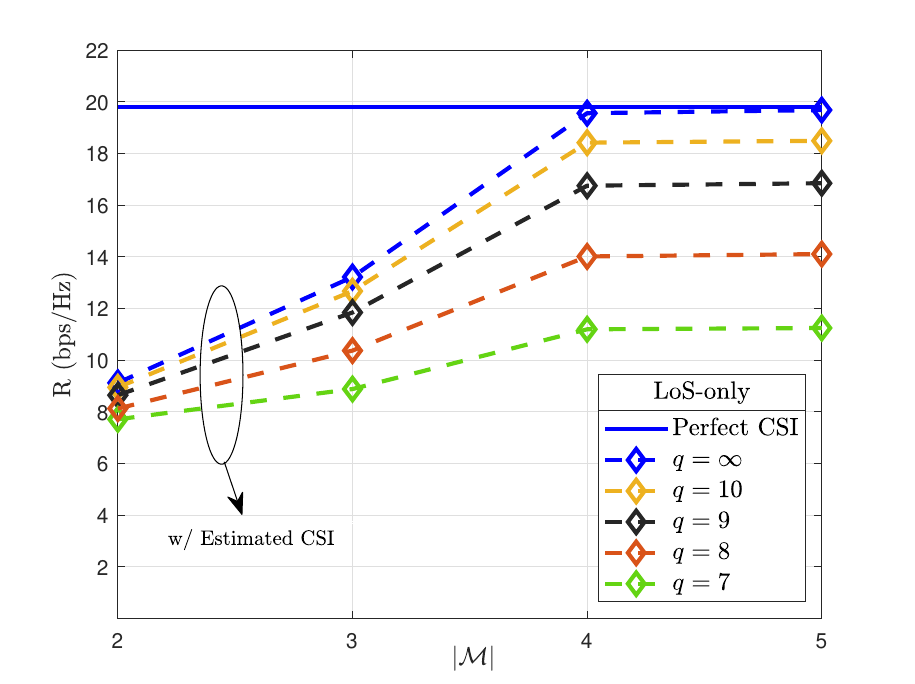}
    \caption{Achievable rate of a UE vs. $|\setM|$ (LoS-only channel setup).}
    %\vspace{-0.5cm}
    \label{fig:rate_pilot}
\end{figure}
%***************************
%******************
%*****************************
\begin{figure}[t]
    \centering
    \begin{subfigure}[b]{1\linewidth}
        \centering       \includegraphics[width=\linewidth]{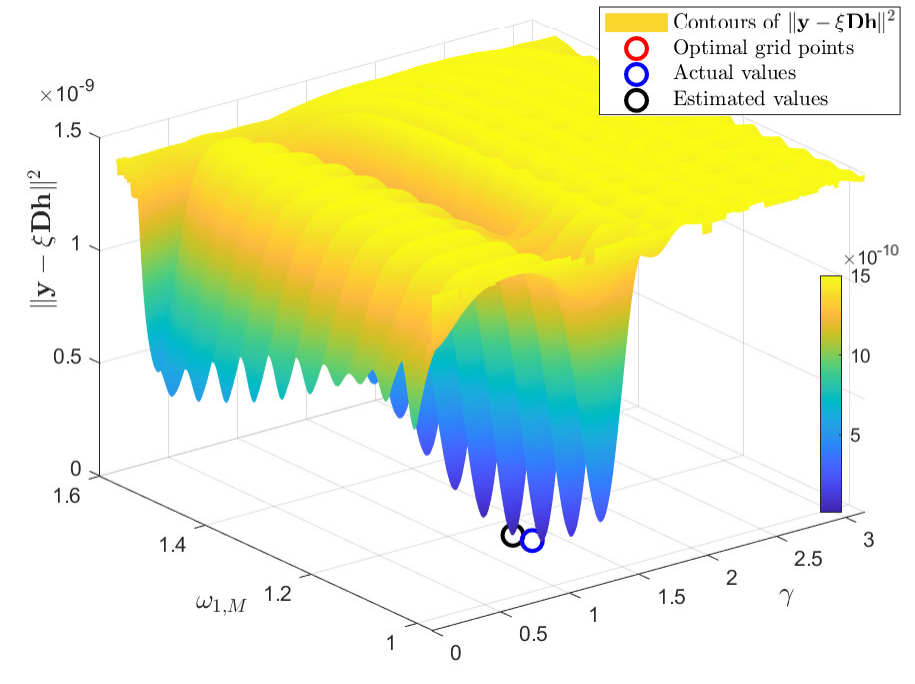}
        \caption{$\|\mathbf{y-\xi\mathbf{D}\mathbf{h}}\|^2$ v.s. $(\omega_{1,1},\omega_{1,M})$ for $|\setM|=2$.}
        \label{fig:liklihood2P}
    \end{subfigure}    
    \begin{subfigure}[b]{1\linewidth}
        \centering        \includegraphics[width=\linewidth]{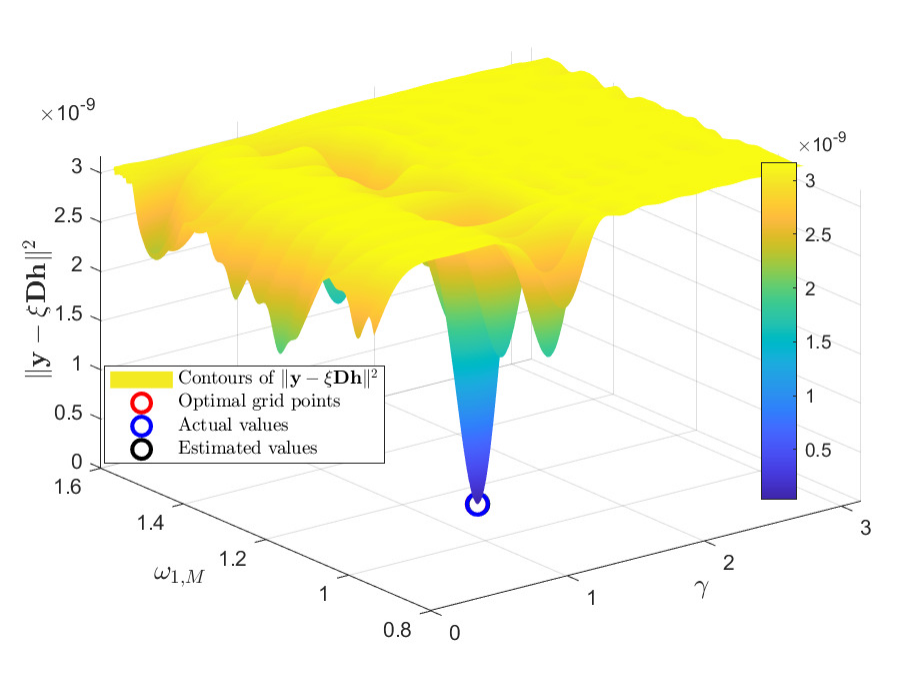}
        \caption{$\|\mathbf{y-\xi\mathbf{D}\mathbf{h}}\|^2$ v.s. $(\omega_{1,1},\omega_{1,M})$ for $|\setM|=4$.}
        \label{fig:liklihood4P}
    \end{subfigure}
    \caption{The objective function of the LoS angles
estimator as a function of $(\omega_{1,M}, \gamma)$, with $\omega_{1,1}=\hat{\omega}_{1,1}$ for a single UE realization.}
    %\vspace{-0.5cm}
    \label{fig:Likelihood}
\end{figure}
%***************************
%**********************
% \begin{figure*}[t]
%     \centering
%     % First subfigure
%     \begin{subfigure}[b]{0.3\textwidth} % Set each subfigure width to 0.3\textwidth
%         \centering
%         \includegraphics[width=1.02\linewidth]{Fig2/omega_gam.eps} % Adjust width to fill the subfigure
%         \caption{CRLB of  $\omega_{1,M}$ v.s. rotation angle for $d_1=5.5 \text{m}$ and $\beta=60, -17.34~\text{(deg)}$}
%         \label{fig:CRB_omega_gamma}
%     \end{subfigure}
%     \hspace{0.01\textwidth} % Adjust horizontal spacing between subfigures
%     % Second subfigure
%     \begin{subfigure}[b]{0.3\textwidth}
%         \centering
%         \includegraphics[width=\linewidth]{Fig2/phi_gam_subplot.eps}
%         \caption{CRLB of $\gamma$ v.s. rotation angle for $d_1=5.5 \text{m}$ and $\beta=60, -17.34~\text{(deg)}$}
%         \label{fig:CRBgam_vs_gamma}
%     \end{subfigure}
%     \hspace{0.01\textwidth} % Adjust horizontal spacing between subfigures
%     % Third subfigure
%     \begin{subfigure}[b]{0.3\textwidth}
%         \centering
%         \includegraphics[width=\linewidth]{Fig2/Peak_Location.eps}
%         \caption{UE position corresponding to the CRLB peaks}
%         \label{fig:Location}
%     \end{subfigure}
%      \hspace{0.01\textwidth} % 
%     \caption{CRLBs of NF LoS reference angles for different UE rotation, assuming perfect quantization.}
%     \label{fig:CRB_gamma}
% \end{figure*}
%************************

We begin by evaluating the MSE of the LoS angles estimates and comparing them with their CRLB in Figure~\ref{fig:MSE_CRLB}, considering a single UE scenario with a plain LoS channel. The results show that with a perfect quantizer (i.e., $q = \infty$), as few as 3 or 4 pilots (i.e., pilot-transmitting BS antennas) are sufficient to accurately estimate the reference angles.
When a quantizer with $q$ bits per angle is applied, we observe that $q=8$ bits can adequately preserve estimation accuracy. Additionally, the estimation error of $\gamma$ is higher than that of $\omega_{1,1}$ and $\omega_{1,M}$, which exhibit similar error ranges. %Lastly, it is observed that the CRLB for the $3$-pilot case performs similarly to that of the $4$-pilot case, while its MSE is closer to that of the $4$-pilot case. %\textbf{ This discrepancy arises because the CRLB does not account for the search space $\boldsymbol{\setA}$,  which can lead to larger errors in practice for the 3-pilot case.}
%To assess the impact of angle quantization errors on the accuracy of the LoS MIMO channel, we compute the achievable rate of the UE v.s. number of pilot transmitting BS anteenas in Figure~\ref{fig:rate_pilot}, by running Algorithm~\ref{alg_5} up to \texttt{S.4} to compute precoding (i.e., before the OTA phase) using the estimated LoS CSI, and performing a single OTA iteration or replacing true channel in~\eqref{eq:u_ks} to compute combining.
To assess the impact of angle quantization errors on LoS MIMO channel accuracy, we plot the achievable rate of the UE versus the number of pilot-transmitting BS antennas in Figure~\ref{fig:rate_pilot}, where precoding is obtained by running Algorithm\ref{alg_5} up to \texttt{S.4} using estimate CSI, and combining is computed via a single OTA iteration. 
%or by substituting the true channel into~\eqref{eq:u_ks}.
It is seen that with $q=\infty$, acquiring LoS CSI with only two pilots results in an achievable rate less than half of that in the perfect CSI case. However, starting from three pilots, the achievable rate improves, with the four-pilot case yielding a rate close to that of perfect CSI. Applying quantization with $q=9$ or $q=10$ bits per reference angle provides performance nearly identical to that of the perfect quantization scenario.
% %**********************
% \begin{figure*}[t]
%     \centering
%     % First subfigure
%     \begin{subfigure}[b]{0.3\textwidth} % Set each subfigure width to 0.3\textwidth
%         \centering
%         \includegraphics[width=1.02\linewidth]{Fig2/omega_gam.eps} % Adjust width to fill the subfigure
%         \caption{CRLB of  $\omega_{1,M}$ v.s. rotation angle for $d_1=5.5 \text{m}$ and $\beta=60, -17.34~\text{(deg)}$}
%         \label{fig:CRB_omega_gamma}
%     \end{subfigure}
%     \hspace{0.01\textwidth} % Adjust horizontal spacing between subfigures
%     % Second subfigure
%     \begin{subfigure}[b]{0.3\textwidth}
%         \centering
%         \includegraphics[width=\linewidth]{Fig2/phi_gam_subplot.eps}
%         \caption{CRLB of $\gamma$ v.s. rotation angle for $d_1=5.5 \text{m}$ and $\beta=60, -17.34~\text{(deg)}$}
%         \label{fig:CRBgam_vs_gamma}
%     \end{subfigure}
%     \hspace{0.01\textwidth} % Adjust horizontal spacing between subfigures
%     % Third subfigure
%     \begin{subfigure}[b]{0.3\textwidth}
%         \centering
%         \includegraphics[width=\linewidth]{Fig2/Peak_Location.eps}
%         \caption{UE position corresponding to the CRLB peaks}
%         \label{fig:Location}
%     \end{subfigure}
%      \hspace{0.01\textwidth} % 
%     \caption{CRLBs of NF LoS reference angles for different UE rotation, assuming perfect quantization.}
%     \label{fig:CRB_gamma}
% \end{figure*}

\begin{figure}[t]
    \centering
    % \begin{subfigure}[b]{1\linewidth}
    %     \centering
        \includegraphics[width=\linewidth]{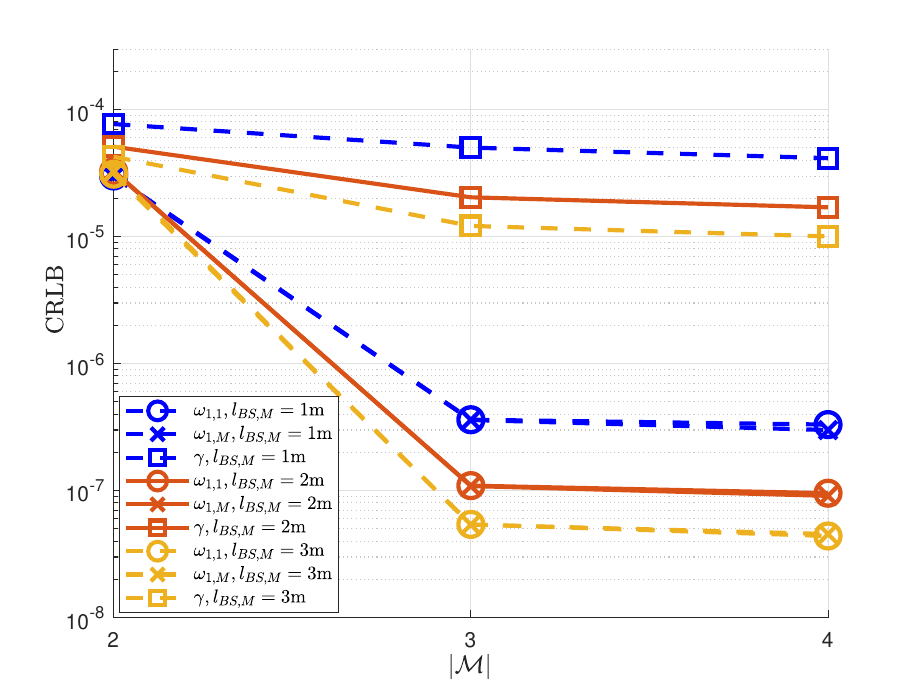}
    %     \caption{CRLB of LoS reference angles vs. $|\setM|$ under Different BS Array Sizes}
    %     \label{fig:CRLB_LUE}
    % \end{subfigure}
    
    % \begin{subfigure}[b]{1\linewidth}
    %     \centering
    %     \includegraphics[width=\linewidth]{Fig2/CRLB_L_ue.eps}
    %     \caption{CRLB of LoS reference angles vs. $|\setM|$ under Different UE Array Sizes}
    %     \label{fig:CRLB_TX}
    % \end{subfigure}
    \caption{Impact of the BS antenna array size on the CRLB of LoS angles for different $|\setM|$.}
    %\vspace{-0.5cm}
    \label{fig:CRLB_pilot_Length}
\end{figure}

\begin{figure}
    \centering
    \begin{subfigure}[b]{1\linewidth}
        \centering        \includegraphics[width=\linewidth]{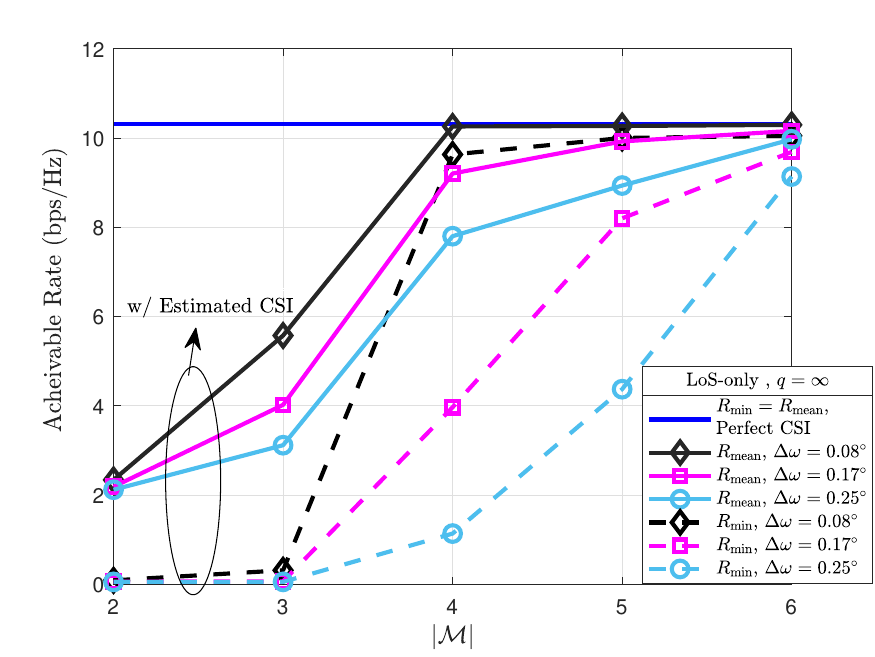}
        \caption{$R_\text{mean}$ and $R_\text{min}$ vs. number of pilot-transmitting BS antennas.}
        \label{fig:rmin_pilot}
    \end{subfigure}    
    \begin{subfigure}[b]{1\linewidth}
        \centering       \includegraphics[width=\linewidth]{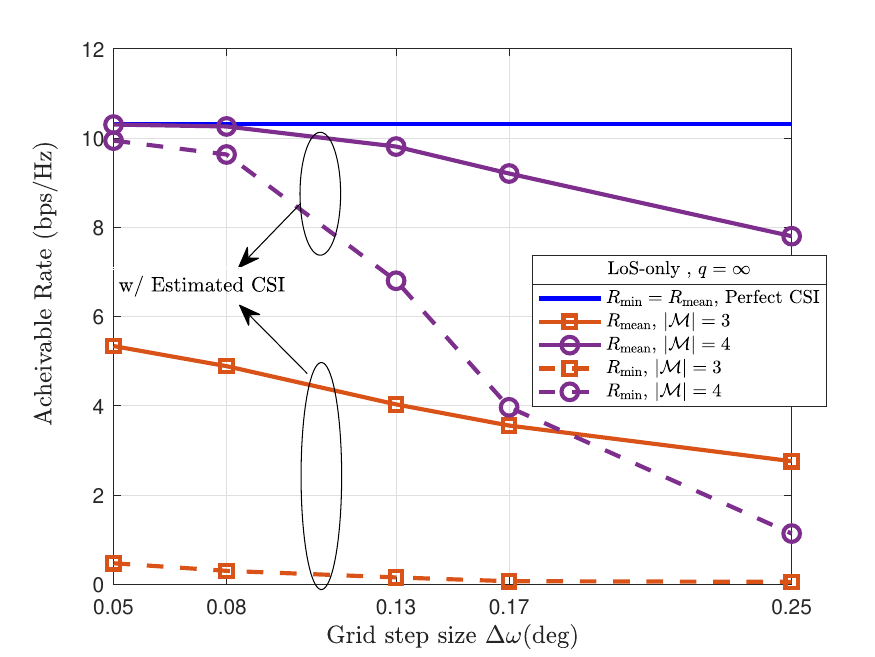}
        \caption{$R_\text{mean}$ and $R_\text{min}$ vs. angular grid step size.}
        \label{fig:rmin_grid}
    \end{subfigure}
    \caption{ Evaluating the accuracy of multi-UE LoS CSI acquisition (without precoding refinement phase), for a LoS-only channel.}
       %\vspace{-0.5cm}
    \label{fig:r_min_Los_only}
\end{figure}

\begin{figure}%[b]{1\linewidth}
        \centering       \includegraphics[width=1\linewidth]{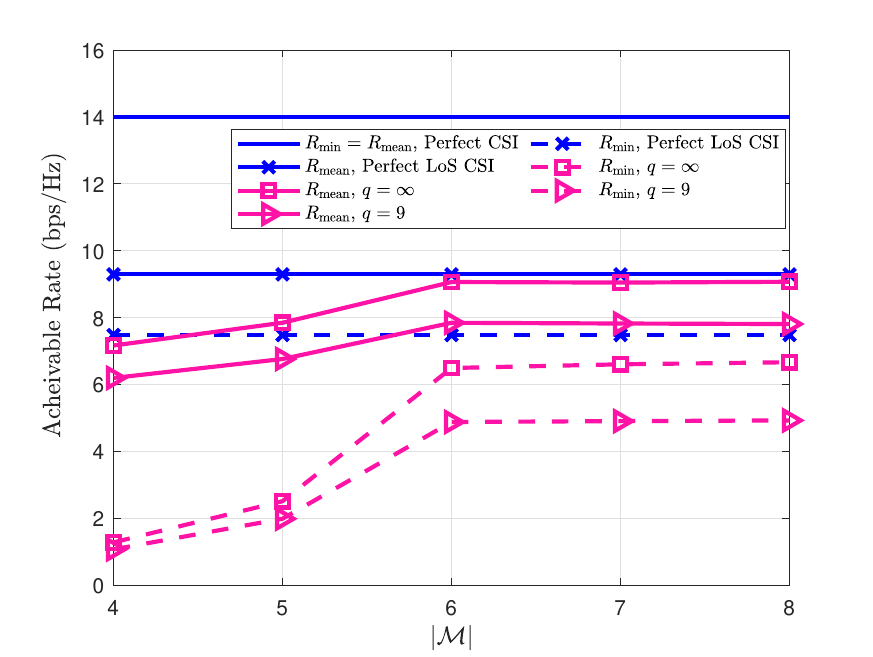}        \caption{$R_\mathrm{mean}$ and $R_\mathrm{min}$ computed without precoding refinement, vs. $|\setM|$. The channel includes both LoS and NLoS components.}           %\vspace{-0.5cm}
        \label{fig:rmin_pilot_Nlos}
    \end{figure}

\begin{figure}%[b]{1\linewidth}
    %\vspace{-0.6cm}
        \centering       \includegraphics[width=1\linewidth]{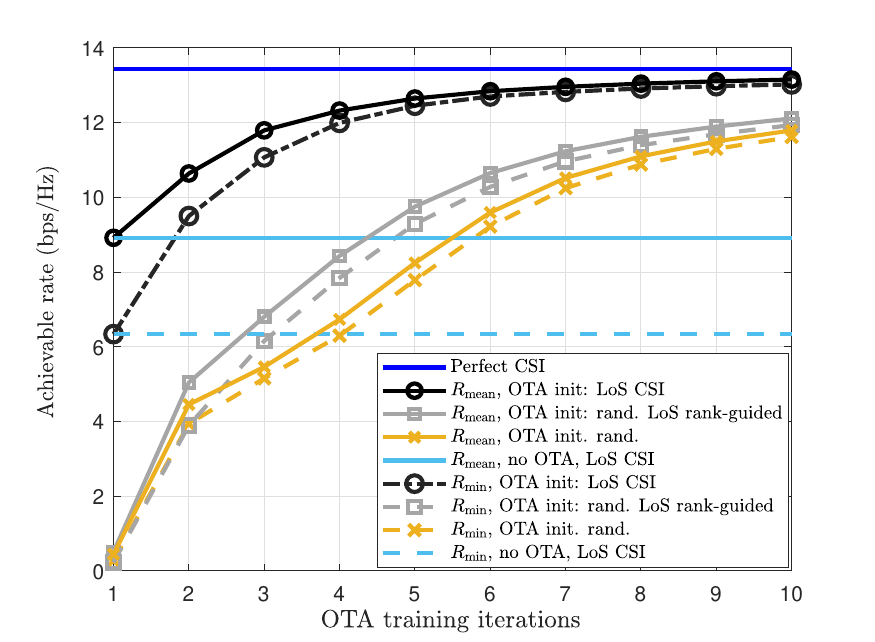}       \caption{$R_\text{mean}$ and $R_\text{min}$ computed after precoding refinement via OTA.
        The channel includes both LoS and NLoS components.}
       %\vspace{-0.6cm}
       % Minimum UE rate comparison of the OTA with the initial
%precoders based on perfect CSI, estimated LoS CSI via 6 pilots,
%estimated LoS channels rank, and No CSI cases.
        \label{fig:BiT}
    \end{figure}

%******************
To visualize the impact of the number of downlink pilots, we plot the utility function of the ML-based parameter estimator, i.e., $\|\mathbf{y}-\xi(\boldsymbol{\omega})\mathbf{D}(\boldsymbol{\omega})\tilde{\mathbf{h}}(\boldsymbol{\omega})\|^2$,  as a function of $(\omega_{1,M},\gamma)$, while fixing $\omega_{1,1}$ to its estimated value $\hat{\omega}_{1,1}$, as shown in Figure~\ref{fig:Likelihood}.
This is done for a single realization of the UE location and its estimated reference angles. The figure demonstrates that increasing the number of pilots from $2$ to $4$ reduces the number of local minima around the optimal point (i.e., the best grid values). With $4$ pilots, only a single dominant peak appears, leading to more accurate angle estimation.
In Figure~\ref{fig:CRLB_pilot_Length}, we evaluate the impact of antenna array length on the accuracy of the parameter estimation algorithm through the CRLB of $(\omega_{1,1},\omega_{1,M},\gamma)$. The results indicate that increasing the BS antenna array lengths reduces angle estimation errors, as it enhances the separability and distinguishability of antenna-pair-specific AoDs. A similar trend is observed with increased UE antenna spacing; the figure is omitted due to space constraints.
%***************************
%*****************************
\begin{figure}[h]
    \centering
    \begin{subfigure}[b]{1\linewidth}
        \centering        \includegraphics[width=\linewidth]{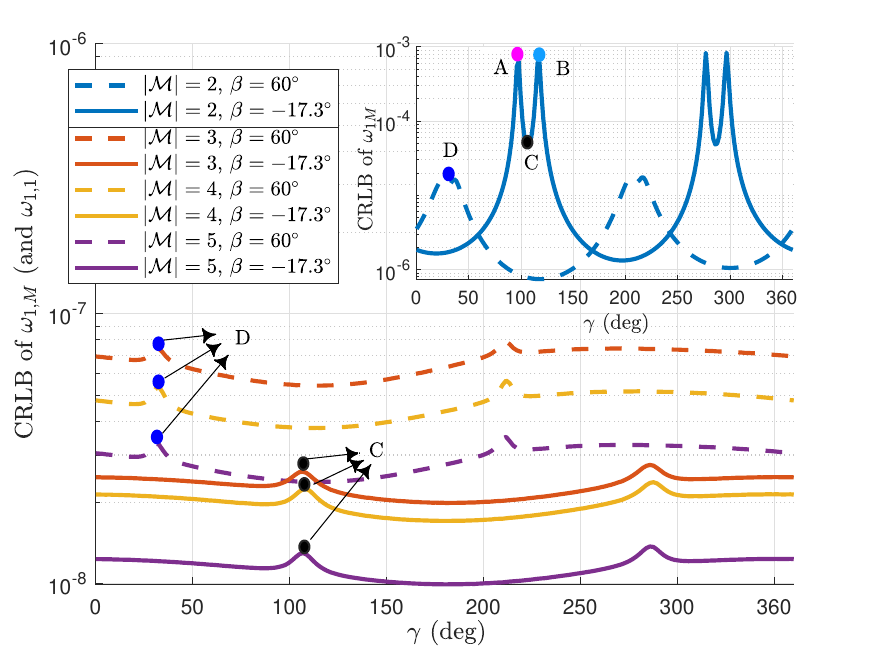}
        \caption{CRLB of $\omega_{1,M}$(and $\omega_{1,1}$) for $d=5.5\text{m}$, and $\beta=60^\circ$ and $-17.3^\circ$.}
        \label{fig:CRLB_omega_vs_gamma}
    \end{subfigure}    
    \begin{subfigure}[b]{1\linewidth}
        \centering       \includegraphics[width=\linewidth]{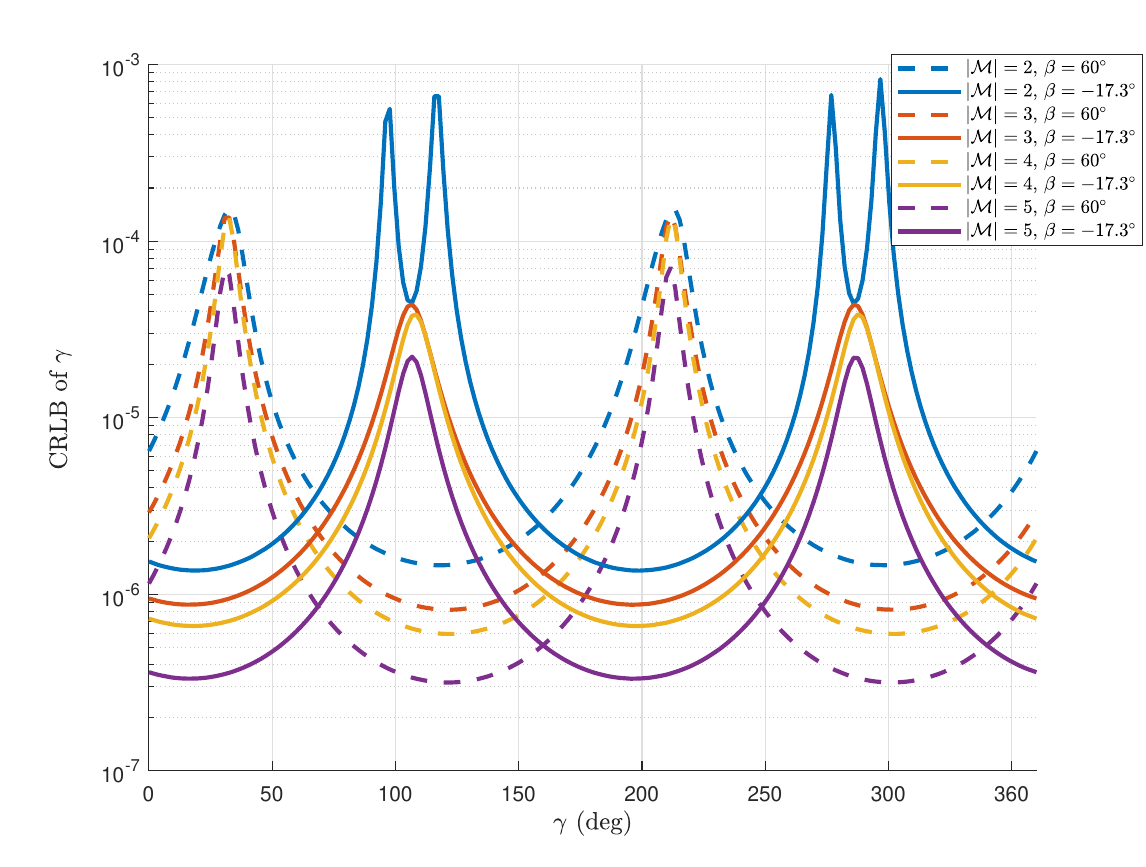}
        \caption{CRLB of $\gamma$ for $d=5.5\text{m}$, and $\beta=60^\circ$ and $-17.3^\circ$.}
        \label{fig:CRLB_gamma_vs_gamma}
    \end{subfigure}
    \caption{CRLBs of NF LoS reference angles for different UE rotation, assuming perfect quantization.}
    \label{fig:CRLB_angle_vs_gamma}
\end{figure}
%***************************
% %*****************************
% \begin{figure}[t]
%     \centering
%     \begin{subfigure}[b]{1\linewidth}
%         \centering        \includegraphics[width=\linewidth]{Fig2/omega_gam.eps}
%         \caption{CRLB of $\omega_{1,M}$(and $\omega_{1,1}$) for $d=5.5\text{m}$, and $\beta=60^\circ$ and $-17.3^\circ$.}
%         \label{fig:CRLB_omega_vs_gamma}
%     \end{subfigure}    
%     \begin{subfigure}[b]{1\linewidth}
%         \centering       \includegraphics[width=\linewidth]{Fig2/phi_gam_subplot.eps}
%         \caption{CRLB of $\gamma$ for $d=5.5\text{m}$, and $\beta=60^\circ$ and $-17.3^\circ$.}
%         \label{fig:CRLB_gamma_vs_gamma}
%     \end{subfigure}
%     \caption{CRLBs of NF LoS reference angles for different UE rotation, assuming perfect quantization.}
%     \label{fig:CRLB_angle_vs_gamma}
% \end{figure}
% %***************************
%****************************
\begin{figure}[t!]
    \centering
    %\vspace{-0.5cm}
    \includegraphics[width=1 \columnwidth]{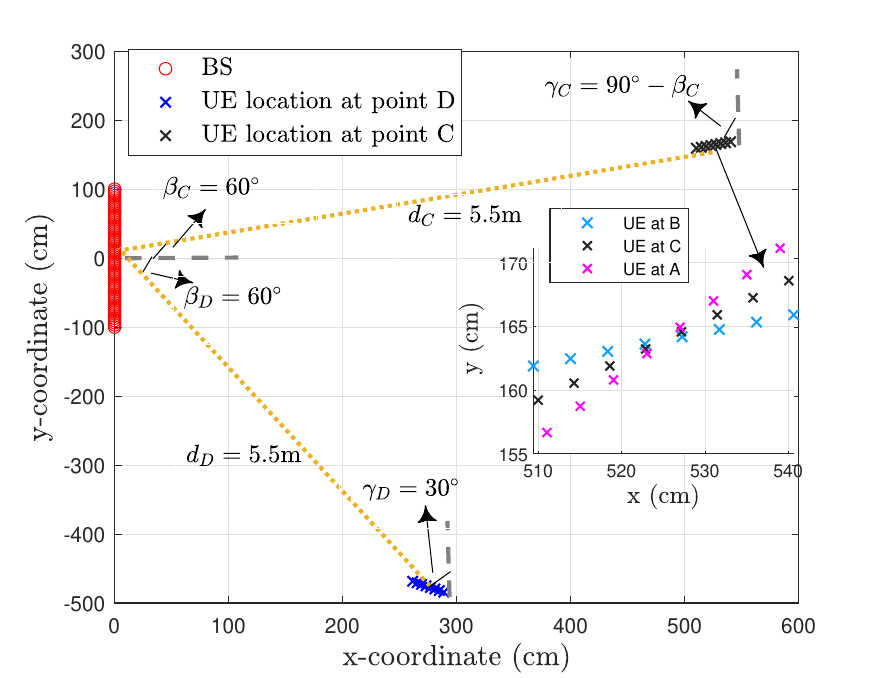}
    \caption{UE position corresponding to the CRLB
peaks of Figure~\ref{fig:CRLB_angle_vs_gamma}.}
    %\vspace{-0.5cm}
    \label{fig:Location}
\end{figure}
%***************************
Figure~\ref{fig:CRLB_angle_vs_gamma} presents the CRLB of reference angles, assuming fixed $d$ and $\beta$ while varying the rotation angle $\gamma$ (Index $k$ omitted for simplicity (single UE case)). Beyond the effect of pilot count, i.e., $|\setM|$, on angles estimation error, we observe that for a given $d$ and $\beta$, a specific $\gamma$ leads to a CRLB peak. For instance, at $d=5.5 \text{m}$ and $\beta=60^\circ$, the worst estimation occurs at $\gamma=90^\circ-\beta=30^\circ$ (i.e., the points D), indicating a poor UE position relative to the BS. %as illustrated in Figure~\ref{fig:Location}.
Similarly, another unfavorable case arises at $d=5.5 \text{m}$, $\beta=-17.34^\circ$, 
 and $\gamma=90^\circ-\beta=107.34^\circ$, (i.e., the points C). %depicted in Figure~\ref{fig:Location} as well.
 Moreover, in the $2$-pilot case, two additional peaks, labeled as points A and B, are observed. Figure~\ref{fig:Location} illustrates the position of the UE at these points, along with point C. The UE locations at points A and B appear equally likely and indistinguishable based on the available geometric information, leading to higher estimation errors. In contrast, at point C, the spatial separability of the UE is greater, reducing uncertainty and improving estimation accuracy.
  While the 2-pilot case exhibits two CRLB peaks, increasing the pilot count reduces the number and intensity of these peaks. 
Figures~\ref{fig:r_min_Los_only} and~\ref{fig:rmin_pilot_Nlos} evaluate the accuracy of the proposed NF LoS MIMO CSI acquisition algorithm in a multi-UE scenario using two evaluation metrics: $R_{\min} = \min\limits_{k} R_k$ and $R_{\mathrm{mean}} = \frac{1}{K}\sum_{k} R_k$ which represent the rate of the worst-performing UE and the average UEs rate, respectively. To compute these rates, the BS independently estimates the LoS CSI for each UE and runs Algorithm~\ref{alg_5} up to step $\texttt{S.4}$ to design the precoding vectors. The combining vectors are obtained by performing a single iteration of the OTA phase.
%or equivalently by substituting the true channel into~\eqref{eq:u_ks}.
%The BS individually estimates the CSI of each UE and executes the Algorithm~\ref{alg_5} up to the step $\texttt{S.4}$ to design precoding vectors. 
%applies the first stage of precoding design based on the estimated LoS channels. 
%The evaluation metric is $R_{\min} = \min\limits_{k} R_k$, representing the worst-performing UE. It is noted that the combining vectors to calculate $R_{\min}$ can be computed by running a single iteration of the OTA phase. 
%the BS independently estimates the LoS CSI for each UE and executes Algorithm~\ref{alg_5} up to step $\texttt{S.4}$ to design coarse precoding vectors. 
%We consider two evaluation metrics defined as $R_{\min} = \min\limits_{k} R_k$ and $R_{\mathrm{mean}} = \frac{1}{K}\sum_{k} R_k$ representing the rate of the worst-performing UE and the mean rate, respectively.
%Notably, the combining vectors required to compute $R_\text{mean}$ and $R_{\min}$ can be obtained by performing a single iteration of the OTA phase.
Figure~\ref{fig:rmin_pilot} shows that 2 or 3 pilots yield poor estimation accuracy in the LoS-only multiuser scenario. However, with 4 pilots, $R_{\min}$ improves significantly and approaches the perfect CSI rate when using a fine grid. Figure~\ref{fig:rmin_grid} further confirms that, even with a fine grid, the 3-pilot case does not provide a sufficient minimum rate, whereas the 4-pilot case offers a wider range of  $R_{\min}$ depending on grid resolution. In particular, $R_{\min}$ is a stringent metric, as it reflects the performance of the worst UE. In contrast, mean-rate metric suggest that 3 pilots may still be viable. For instance, the mean-rate for the 3-pilot case with $\Delta\omega=0.08^\circ$. is $5.57~\text{bps/Hz}$ compared to the mean-rate of the perfect CSI case, which is $10.32~\text{bps/Hz}$.
%mean-rate or sum-rate metrics suggest that 2 or 3 pilots may still be viable. 
%shows that in the presence of significant NLoS channel components, 4 pilots are no longer sufficient for $R_\text{min}$, requiring additional pilots to approach the perfect LoS CSI rate.
Figure~\ref{fig:rmin_pilot_Nlos} shows that in the presence of significant NLoS channel components, using only 4 pilots is no longer sufficient when considering the $R_\text{min}$ metric. Additional pilots are required to approach the performance achieved with perfect LoS CSI.
Finally, in Figure~\ref{fig:BiT}, we evaluate the proposed two-stage precoding design in the presence of significant NLoS channel components in terms of $R_\text{min}$ and $R_\text{mean}$ computed after each OTA iteration. 
Carrying out OTA refinement using initial precoding vectors designed from the estimated LoS CSI achieves near-optimal rates within a few iterations, significantly outperforming the traditional OTA procedure initialized with random precoding with no CSI used during initialization.
Alternatively, to reduce feedback overhead, each UE could report only the rank of its LoS channel instead of the full set of reference angles. Based on this rank information, the BS assigns streams accordingly and applies random precoding for each stream. However, this rank-guided initialization provides only limited performance gains compared to the OTA approach with random initialization not guided by any form of CSI.
% %********************
%*****************************
% \begin{figure}[t]
%     \centering
%-------------------------------
\section{Conclusions}
\label{sec:CONC}
%conclusion
A geometry-aided NF LoS CSI acquisition scheme at the BS was proposed in this paper, followed by a two-stage transmission design procedure. The NF LoS MIMO channel was characterized using three angular parameters, %replacing the conventional reference-distance-based approach commonly employed in existing schemes.
improving the conventional reference-distance-based NF channel parametrization approaches commonly employed in existing schemes.
A limited number of downlink pilots were used to enable the estimation of LoS angular parameters at each UE. 
%The proposed method was shown to enable highly accurate multi-UE LoS channel reconstruction using as few as 4–6 pilot sequences
%The proposed method demonstrated highly accurate multi-UE LoS channel reconstruction with only 4–6 pilot sequences
The proposed method was shown through simulations and CRLB analysis to achieve highly accurate multi-UE LoS channel reconstruction using as few as 4–6 pilot sequences—significantly fewer than the number of BS antennas, $M$ (i.e., $4$–$6 \ll M$). In the first stage of the proposed transmission scheme, the BS collects LoS CSI based on angular triplets estimated and fed back by the UEs. A coarse precoding design is then carried out at the BS based on the estimated multi-UE LoS MIMO CSI. This is followed by a refinement stage involving OTA training iterations between the BS and UEs. The proposed two-stage design was shown to converge to the full-CSI performance within only a few OTA iterations, significantly outperforming conventional OTA procedures initialized with random precoding.
%\vspace{-2mm}
%     \caption{Performance of the proposed transmission design.}
%         \vspace{-0.5cm}
%     \label{fig:rate-BiT}
% \end{figure}
%********************
%**********************************
\appendix
\section{}
\subsection{ Precoding Design} \label{app:A3}
The symmetric rate maximization problem is formulated as 
%--------------------------
 \begin{equation}
\label{eq:Optimization_problem}
\begin{aligned}
 \max_{\u_{k,s}, \m_{k,s}} & \min_{k \in \setI_K} \sum_{s=1}^{S_k} \log(1+\hat{\Gamma}_{k,s}) \\ \text{s. t.} & \quad \sum_{k=1}^K\sum_{s=1}^{S_k} \|\m_{k,s}\|^2 \leq P_\text{BS},
\end{aligned}
\end{equation}
%---------------------------
and it is not jointly convex in $\mathbf{u}_{k,s}$ and $\mathbf{m}_{k,s}$.  For fixed $\mathbf{m}_{k,s}, \forall (k,s)$, the rate-maximizing combining are given by~\eqref{eq:u_ks}.
For fixed $\mathbf{u}_{k,s}, \forall (k,s)$, the problem can be solved iteratively following~\cite [Sections \MakeUppercase{\romannumeral 3}-\MakeUppercase{\romannumeral 4}]{Jarkko}. 
% . From the KKT conditions, the closed-form precoding solution is given by~\eqref{eq:m_kl}.
%*********************************
%To convexify~\eqref{eq:Optimization_problem}
%Using the MSE-SINR relation in \eqref{eq:Optimization_problem}, the problem is reformulated in terms of MSE estimates $\hat{\epsilon}_{k,s}$ of the $s^\text{th}$ stream of UE $k$ as~\cite{Jarkko}
The MSE estimates of the $s^\text{th}$ stream of UE $k$ is given as~\cite{Jarkko}
%********************
    \begin{align}\label{eq:mse}    \hat{\epsilon}_{k,s}&=\Exp\Big[\big|\u_{k,s}^\herm {\mathbf{y}}_k-x_{k,s}\big|^2\Big]=\big|1-\u_{k,s}^\herm \hat{\H}_k \m_{ k, s}\big|^2\\  &+\sum\limits_{\bar k=1}^K \sum\limits_{\substack{\bar s=1 
 \\ (\bar{k}, \bar{s}) \neq (k, s)}}^{S_{\bar k}} \big|1-\u_{k,s}^\herm \hat{\H}_k \m_{ \bar k, \bar s}\big|^2
+\sigma_{\text{UE}_k}^2\|\u_{k,s}\|^2.\nonumber
\end{align}
%*********************
% where $\hat{\mathbf{x}}_k$ is obtained by replacing $\mathbf{H}_k$ by $\hat{\mathbf{H}}_k$ in \eqref{eq:y_k}.
%The combining vector~\eqref{eq:u_ks} is the linear minimum-MSE (MMSE) receiver. 
 For a fixed  $\mathbf{u}_{k,s}$, \eqref{eq:mse} is convex in $\mathbf{m}_{k,s}  \forall (k,s)$, and vise versa, and the optimal combining minimizing~\eqref{eq:mse} is given by \eqref{eq:u_ks}.
 %Plugging optimal combining~\eqref{eq:u_ks} into~\eqref{eq:mse}, we have $\hat{\epsilon}_{k,s}=(1+\hat{\Gamma}_{k,s})^{-1}$~\cite{Jarkko}. Using the MSE-SINR relation in \eqref{eq:Optimization_problem}, the problem is reformulated in terms of $\hat{\epsilon}_{k,s}$.
 Substituting the optimal combining from~\eqref{eq:u_ks} into~\eqref{eq:mse} yields $\hat{\epsilon}_{k,s}=(1+\hat{\Gamma}_{k,s})^{-1}$~\cite{Jarkko}, based on which~\eqref{eq:Optimization_problem} is reformulated in terms of $\hat{\epsilon}_{k,s}$.
However, with fixed $\mathbf{u}_{k,s}$, \eqref{eq:Optimization_problem} is still non-convex w.r.t  $\mathbf{m}_{k,s}, \forall (k,s)$. 
 Following~\cite{Jarkko}, by introducing the auxiliary constraint $\hat{\epsilon}_{k,s} \leq 2^{t_{k,s}}$ and applying a first-order Taylor approximation to the non-convex constraint,~\eqref{eq:Optimization_problem} is approximated as  %reformulated as a difference of convex functions program (DCP). Successive convex approximation (SCA) is applied to bound the non-convex constraints, yielding an approximated convex optimization problem
%*******************************************
% \begin{align} \label{eq:cvx_problem}
% \begin{array}{clclc}
% \underset{\m_{k,s}, t_{k,s}, r_c}{\text{max}}  & r_c & \\
% \text{s.t.} & & \\
% \lambda_k: & r_c \leq \sum_{s \in \setS_k} t_{k,s} ,  \forall k  \\
%   \nu_{k,s}: &\hat{\epsilon}_{k,s} \leq -\Bar{a}_{k,s}t_{k,s}+ \Bar{b}_{k,s}, & \forall (k,s) \\
%  \zeta:& \sum_{k \in \setK}\sum_{s \in \setS_k}  \|\v_{k,s}\|^2 \leq P_\text{BS} &
% \end{array}
% \end{align}
%*******************************************
 \begin{align} \label{eq:cvx_problem222}
\underset{\m_{k,s}, t_{k,s}, r_c}{\text{max}} & \quad r_c\nonumber \\
\text{s.t.} \quad & \lambda_k: \quad r_c \leq \sum_{s=1}^{S_k} t_{k,s}, \quad \forall k  \nonumber\\ & \nu_{k,s}: \quad \hat{\epsilon}_{k,s} \leq -\Bar{\alpha}_{k,s} t_{k,s} + \Bar{\beta}_{k,s}, \quad \forall (k,s) \nonumber\\ & \eta: \quad \sum_{k=1}^{K}\sum_{s=1}^{S_k}  \|\m_{k,s}\|^2 \leq P_\text{BS},
\end{align}
%*******************************************
where $r_c$ is the common rate introduced while rewriting \eqref{eq:Optimization_problem} in epigraph form, and $\lambda_k$, $\nu_{k,s}$, and $\zeta$ are the dual variables. %corresponding to each constraint. 
The approximation points are given as~\cite{Jarkko}
\begin{equation}
\label{eq:a_b}
    \begin{aligned}    \Bar{\alpha}_{k,s}=2^{\Bar{t}_{k,s}}\log_e2, \quad  \Bar{\beta}_{k,s}=2^{\Bar{t}_{k,s}}(1+\Bar{t}_{k,s}\log_e2), 
\end{aligned}
\end{equation}
where $\Bar{t}_{k,s}$ is the solution from the previous iteration. Algorithm~\ref{alg_3} summarizes the CVX-based solution of \eqref{eq:cvx_problem222}, which can be executed in \texttt{S.3} of Algorithm~\ref{alg_5}.
%\vspace{-1.5cm}
\begin{algorithm}
%\begin{algorithmic}
%\color{gray}
  \textbf{Data:} $\{\hat{\mathbf{H}}_k\}_{k=1}^K$, $\{S_k\}_{k=1}^K$, $\{\sigma^2_{\text{UE}_k}\}_{k=1}^K$, $P_\text{BS}$, $\{\{\hat{\mathbf{u}}_k\}_{s=1}^{S_k}\}_{k=1}^K$.
  
 % \hspace{8mm}$\{l_{\text{TX},m}\}_{j=1}^M$ and $\{l_{\text{RX},i}\}_{i=1}^N$, respectively.
 \vspace{1mm}
 \begin{itemize}[leftmargin=12mm]
%*****************************
\item[\texttt{(S.1)}]
Calculate $\bar{t}_{k,s}  \forall(k,s)$ from $\bar{t}_{k,s}=-\log_2(\bar{\epsilon}_{k,s})$, where $\bar{\epsilon}_{k,s}$ is obtained from~\eqref{eq:mse}  .
%*****************************
\item[\texttt{(S.2)}]
Calculate $\bar{\alpha}_{k,s} \forall(k,s)$  and $\bar{\beta}_{k,s} \forall(k,s)$ from \eqref{eq:a_b}.
%*****************************
\item[\texttt{(S.3)}]
Solve \eqref{eq:cvx_problem222} using CVX and obtain $\mathbf{m}_{k,s} \forall(k,s)$. Note that $\hat{\epsilon}_{k,s}$ is a function of $\{\{\mathbf{m}_{k,s}\}_{s=1}^{S_k}\}_{k=1}^K$ for a fixed $\{\{\mathbf{u}_{k,s}\}_{s=1}^{S_k}\}_{k=1}^K$. 
 \end{itemize}
 %\end{algorithmic}
\caption{CVX-based solution for \eqref{eq:Optimization_problem}, based on~\cite{Jarkko}} \label{alg_3}
\end{algorithm}
%CVX solution of \eqref{eq:Optimization_problem}
%_____________________________________
%-----------------------
%----------------------
%\vspace{-1cm}
\subsection{KKT Solution}
The Lagrangian for~\eqref{eq:cvx_problem222} is given by
%................................
\begin{align}    \mathcal{L}_{\eqref{eq:cvx_problem222}}&=-r_c+\eta \bigg(\sum_{k=1}^{K}\sum_{s=1}^{S_k}  \|\m_{k,s}\|^2-P_\text{BS}\bigg) + \sum_{k=1}^{K}\bigg(\sum_{s=1}^{S_k} \nu_{k,s} \nonumber \\ 
   &\phantom{=} \times \big(\hat{\epsilon}_{k,s} +  \Bar{\alpha}_{k,s} t_{k,s} \! - \! \Bar{\beta}_{k,s}\big) \!+\! \lambda_k \big(r_c -\sum_{s=1}^{S_k} t_{k,s}\big)\bigg),
\end{align}
%-----------------------------------
%where $ \mathcal{L}(.)= \mathcal{L}(\lambda_k,\nu_{k,s},\eta, r_c, \m_{k,s}, t_{k,s})$. 
For a fixed $\u_{k,s}$, $\Bar{\alpha}_{k,s}$ and $\Bar{\beta}_{k,s}$, the KKT conditions are
%the primal and dual variables must satisfy the KKT conditions at the optimal point:
\begin{align}
    \sum_{k=1}^{K} \lambda_k= 1; \quad \nu_{k,s}= \frac{\lambda_k}{\Bar{\alpha}_{k,s}}, \quad \forall (k,s);\label{kkt2}
\end{align}
obtained from $\nabla_{r_c}\mathcal{L}_{\eqref{eq:cvx_problem222}}=0$ and $\nabla_{t_{k,s}}\mathcal{L}_{\eqref{eq:cvx_problem222}}=0$, respectively.
the closed-form precoding solution is obtained from $\nabla_{\nu_{k,s}}\mathcal{L}_{\eqref{eq:cvx_problem222}}=0$ as in~\eqref{eq:m_kl}. Complementary slackness on the second constraint of~\eqref{eq:cvx_problem222} results in $t_{k,s}= {\bar{\alpha}^{-1}_{k,s}}({\bar{\beta}_{k,s} - \hat{\epsilon}_{k,s}}), \forall (k,s)$
% \begin{align}
%     t_{k,s}= \frac{\Bar{\beta}_{k,s} - \hat{\epsilon}_{k,s}}{\Bar{\alpha}_{k,s}}, \quad \forall (k,s) \label{kkt3}
% \end{align}
and the common rate $r_c$ is calculated as  $r_c=\mathop{\text{mean}}\limits_{k}\big(\sum_{s=1}^{S_k} t_{k,s}\big)$.
%\begin{align}    r_c=\mathop{\text{mean}}\limits_{k}\big(\sum_{s=1}^{S_k} t_{k,s}\big).
% \label{rc}
% \end{align}
Finally, $\lambda_k$ is updated by the sub-gradient method as $\lambda_k\leftarrow \text{max} \big(\lambda_k+\mu\nabla_{\lambda_k}\mathcal{L}_{\eqref{eq:cvx_problem222}},0\big)$.
For the KKT-based solution of~\eqref{eq:cvx_problem222}, $\Bar{\alpha}_{k,s}$ and $\Bar{\beta}_{k,s}$ are first computed. Then, the primal and dual variables are updated using the equations above following~\cite{Nariman}.

\end{document}